\PassOptionsToPackage{table}{xcolor}
\documentclass[sigconf,nonacm,screen]{acmart}

\AtBeginDocument{%
}

\setcopyright{acmlicensed}
\copyrightyear{2026}
\acmYear{2026}
\acmDOI{}
\acmConference[ICAIF '26]
  {ACM International Conference on AI in Finance}
  {2026}
  {TBD}
\acmISBN{}

\usepackage{graphicx}
\usepackage{booktabs}
\usepackage{multirow}
\usepackage{array}
\usepackage{tabularx}
\usepackage{makecell}
\usepackage{threeparttable}
\usepackage{amsmath}
\usepackage{enumitem}
\usepackage{pifont}
\usepackage{xspace}
\usepackage{microtype}
\usepackage{placeins}
\usepackage[most]{tcolorbox}
\usepackage{listings}
\usepackage[ruled,vlined,linesnumbered]{algorithm2e}


\definecolor{GainOne}{HTML}{F3F7FB}
\definecolor{GainTwo}{HTML}{E4EEF7}
\definecolor{GainThree}{HTML}{CDDFEF}

\definecolor{LossOne}{HTML}{FCF6F4}
\definecolor{LossTwo}{HTML}{F6E7E3}
\definecolor{LossThree}{HTML}{ECD1CA}

\definecolor{BlockGray}{HTML}{EEF2F5}
\definecolor{BlockText}{HTML}{34424F}
\definecolor{HeaderGray}{HTML}{F7F8FA}
\definecolor{MethodAccent}{HTML}{245A78}
\definecolor{TableRule}{HTML}{AAB4BD}

\newcommand{\gainone}[1]{%
  \cellcolor{GainOne}\rule{0pt}{2.20ex}#1%
}
\newcommand{\gaintwo}[1]{%
  \cellcolor{GainTwo}\rule{0pt}{2.20ex}#1%
}
\newcommand{\gainthree}[1]{%
  \cellcolor{GainThree}\rule{0pt}{2.20ex}#1%
}

\newcommand{\lossone}[1]{%
  \cellcolor{LossOne}\rule{0pt}{2.20ex}#1%
}
\newcommand{\losstwo}[1]{%
  \cellcolor{LossTwo}\rule{0pt}{2.20ex}#1%
}
\newcommand{\lossthree}[1]{%
  \cellcolor{LossThree}\rule{0pt}{2.20ex}#1%
}

\newcommand{\blockrow}[1]{%
  \rowcolor{BlockGray}%
  \multicolumn{6}{@{}l}{%
    \rule{0pt}{2.35ex}%
    \textcolor{BlockText}{\textbf{\strut #1}}%
  }%
}

\newcommand{\method}{\textsc{VolRouter}\xspace}

\newcommand{\methodlabel}{%
  \textcolor{MethodAccent}{\textbf{\method}}%
}

\newcommand{\tableheader}[1]{%
  \cellcolor{HeaderGray}\textbf{#1}%
}

\definecolor{CitationBlue}{HTML}{5F93B7}
\definecolor{SectionRefRed}{HTML}{B76F6F}
\hypersetup{
  colorlinks=true,
  citecolor=CitationBlue,
  urlcolor=CitationBlue,
  linkcolor=black
}
\AtBeginDocument{\hypersetup{citecolor=CitationBlue,urlcolor=CitationBlue,linkcolor=black}}
\newcommand{\SectionRef}[1]{%
  \hyperref[#1]{\textcolor{SectionRefRed}{Section~\ref*{#1}}}%
}
\newcommand{\AppendixRef}[1]{%
  \hyperref[#1]{\textcolor{SectionRefRed}{Appendix~\ref*{#1}}}%
}

\begin{document}

\title{Beyond Forecasting: Recasting Volatility Control as a Routing Problem}

\author{Hongji Pu}
\authornote{Corresponding author.}
\email{hongjip2@illinois.edu}
\affiliation{%
  \institution{University of Illinois, Urbana-Champaign}
  \city{Urbana}
  \state{Illinois}
  \country{USA}
}

\author{Leyang Zhou}
\email{leyangz3@illinois.edu}
\affiliation{%
  \institution{University of Illinois, Urbana-Champaign}
  \city{Urbana}
  \state{Illinois}
  \country{USA}
}

\renewcommand{\shortauthors}{Pu et al.}

\begin{abstract}
Volatility control converts risk estimates into portfolio exposure, yet
many existing approaches rely on a fixed estimator or a pre-specified
control rule across changing market conditions.
We propose \method, a modular framework that formulates volatility control
as state-conditioned routing over estimator--controller pairs.
\method summarizes the current market environment into a control-relevant
state profile and separates routing into state inference, switch review,
and pair selection.
The Router can be instantiated by rule-based, learnable, or LLM-based
decision modules, while portfolio actions remain generated by predefined
control policies. We evaluate \method on S\&P 500, Multi-Asset, Bitcoin, and USDT
volatility-control settings.
\method achieves the highest Sharpe ratio in three of the four settings.
On S\&P 500, it increases Sharpe from 0.952 for RV + Naive Scaling to
1.222 while reducing maximum drawdown from 15.10\% to 12.58\% and
daily CVaR from 1.76\% to 1.32\%.
On Multi-Asset, it improves Sharpe from 1.498 to 1.540 and reduces CVaR
from 1.56\% to 1.18\%.
Bitcoin shows a similar improvement in risk-adjusted performance, whereas
USDT provides a boundary case in which simpler state-aware selectors remain
competitive.
Ablations and sensitivity analyses further indicate that the gains depend
on relative policy evaluation and selective, persistent switching rather
than simply enlarging the policy library.
Overall, the results support viewing volatility control as a policy-selection
problem in settings where control requirements vary across market states.
\end{abstract}

\ccsdesc[500]{Computing methodologies~Machine learning}
\ccsdesc[300]{Applied computing~Economics}

\keywords{volatility targeting, routing, risk control, portfolio management}

\maketitle

\section{Introduction}
\label{sec:introduction}

Volatility control turns a risk estimate into a portfolio exposure
\citep{Markowitz1952,FlemingKirbyOstdiek2001,MoreiraMuir2017,MSCI2021}.
Forecasting risk, however, is only part of the decision. The same increase in
estimated volatility can require different actions in different market states.
At the start of a sharp drawdown, fast de-risking may be useful. During a noisy
reversal, the same reaction can create unnecessary turnover. In a quiet market,
keeping more exposure may be preferable. Thus, a volatility-control system must
answer two questions: \emph{what risk is present}, and \emph{which control rule
should act on that risk now}. Risk-management practice has long recognized this
need for judgment beyond a single mechanical statistic
\citep{LongerstaeyMore1996,AlmgrenChriss2001}.

Most existing work emphasizes one side of this pipeline. Volatility models aim
to improve the risk estimate using realized volatility
\citep{BarndorffNielsenShephard2002,AndersenBollerslevDieboldLabys2003}, EWMA
\citep{LongerstaeyMore1996}, ARCH/GARCH-family models
\citep{Engle1982,Bollerslev1986,Nelson1991,HansenLunde2005}, downside-risk
measures \citep{BarndorffNielsenKinnebrockShephard2008,WangYan2021}, or implied
volatility such as VIX \citep{Whaley2009,Bozovic2024}. A second line of work
studies how a chosen signal is converted into exposure, for example through
target-volatility scaling \citep{MoreiraMuir2017}, index-level risk-control
rules \citep{MSCI2021,SPDJI2025RC2}, drawdown-aware control
\citep{NystrupBoydLindstromMadsen2019}, or regime-aware allocation
\citep{Hamilton1989,AngBekaert2002}. These approaches are useful, but they
usually keep the estimator--controller design fixed once the strategy is
specified. They do not make the choice among multiple valid control policies an
explicit decision at each state.

\begin{figure*}[t]
\centering
    \includegraphics[width=\textwidth]{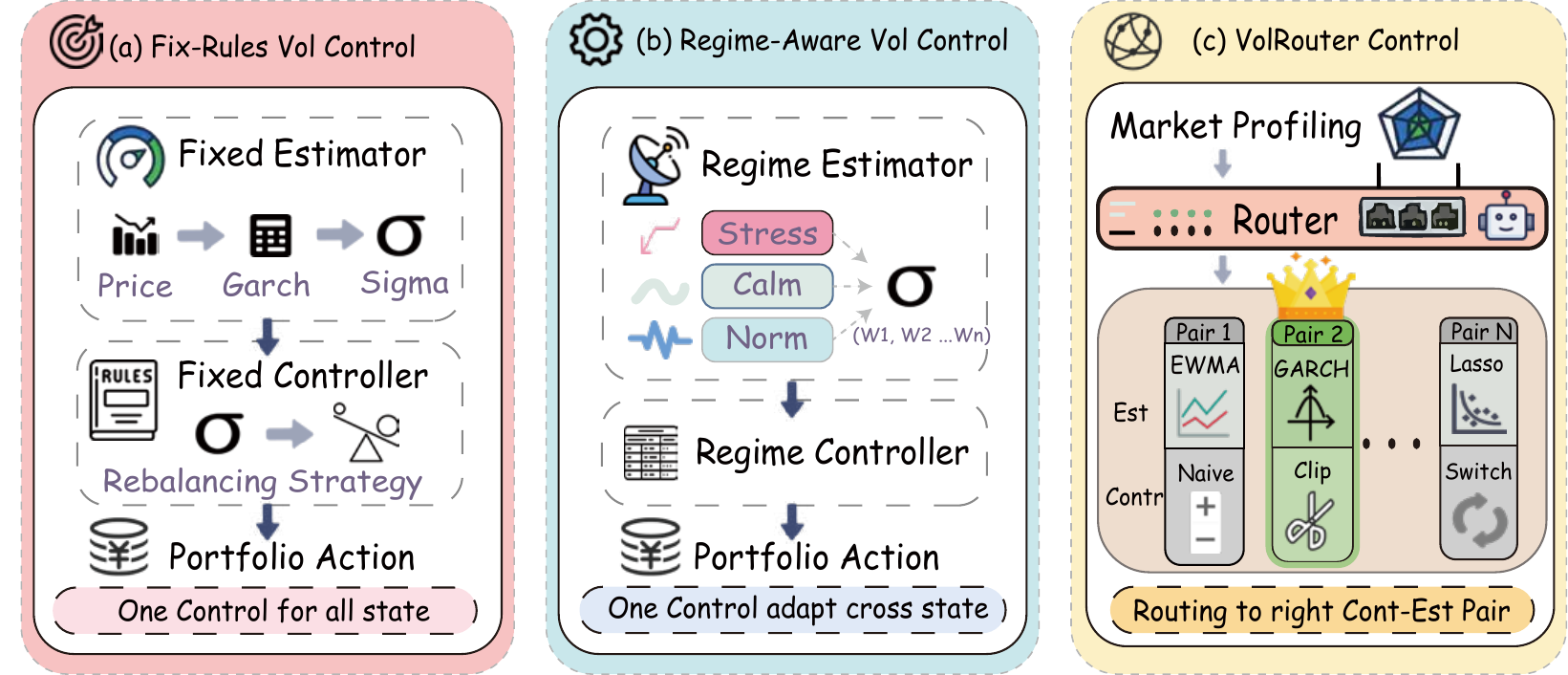}
    \Description{Three-panel schematic comparing fixed-rule volatility control, regime-aware control, and VolRouter. VolRouter adds a state-profiling and policy-routing layer before execution.}
\caption{
Comparison of three volatility-control paradigms.
\textbf{(a) Fixed-rule control} uses one estimator--controller pair across states.
\textbf{(b) Regime-aware control} adapts a fixed control design using state-dependent rules.
\textbf{(c) VolRouter} makes policy selection explicit: it profiles the current state, decides whether the active pair should be kept, and selects a replacement only when switching is justified.
Estimator and controller labels are illustrative; experiment-specific eligibility is defined separately.
}
\label{fig:volrouter_overview}
\end{figure*}

Figure~\ref{fig:volrouter_overview} highlights the missing layer. We represent a
control policy as an \emph{estimator--controller pair}: the estimator measures
risk, and the controller converts that risk object into a portfolio action. For
example, an EWMA estimator can be paired with inverse-volatility scaling, while
a different pair can combine a richer volatility estimate with a drawdown-aware
controller. The question is no longer only which estimator is most accurate.
The system must decide whether the currently active pair remains appropriate and,
if not, which alternative should take control.

We formulate this decision as \textbf{persistent, state-conditioned routing}
and propose \method. The framework first summarizes the current market into a
control-relevant state, then performs three simple decisions: \emph{state
inference}, \emph{switch review}, and \emph{pair selection}. The switch review
is important because changing policies every time rankings move slightly can
create unstable behavior and extra turnover. A replacement is therefore chosen
only after the system decides that the active pair should not be retained.

The Router is a high-level decision module rather than a trading policy. It can
be rule-based, learnable, or LLM-based. In the LLM instantiation studied here,
the language model acts like a constrained supervisory agent: it reads
structured, historical state and candidate diagnostics, returns a state label,
decides \texttt{hold} or \texttt{switch}, and, when needed, selects one valid
pair from a closed candidate list. It never predicts the next return or directly
generates portfolio weights. Tool-routing and agent systems motivate this use of
an LLM as an orchestrator rather than as the downstream executor
\citep{ShenEtAl2023HuggingGPT,WuEtAl2023AutoGen,YaoEtAl2023ReAct}.

Empirically, \method achieves the highest Sharpe ratio in three of four reported
settings. On S\&P 500 and Multi-Asset, the main effect is better risk allocation:
Sharpe improves while downside risk falls, even when raw return does not increase.
Bitcoin shows a similar improvement in risk-adjusted performance. USDT is a
useful boundary case, where simpler state-aware selectors remain competitive.
Ablations further show that the benefit depends on relative candidate evaluation,
temporal context, and selective switching rather than on merely enlarging the
policy library.

Our contributions are threefold. \textbf{First}, we make policy selection an
explicit layer between risk estimation and portfolio execution. \textbf{Second},
we introduce a modular routing framework that separates market interpretation,
the decision to switch, and replacement-pair selection. \textbf{Third}, we
characterize when this routing layer helps through four markets, component
ablations, sensitivity studies, and multiple Router backbones.

\SectionRef{sec:related_work} positions the formulation relative to volatility
control and routing. \SectionRef{sec:methodology} defines the three-stage Router
and its LLM instantiation. \SectionRef{sec:experimental_setup} describes the
comparisons and evaluation protocol, and \SectionRef{sec:results} reports the
main empirical results.

\section{Related Work}
\label{sec:related_work}

\textbf{Traditional volatility control.}
One line of work improves the risk estimate using realized volatility, EWMA,
ARCH/GARCH-family models, downside-risk measures, and implied-volatility signals
\citep{LongerstaeyMore1996,Engle1982,Bollerslev1986,Nelson1991,BarndorffNielsenShephard2002,AndersenBollerslevDieboldLabys2003,HansenLunde2005,WangYan2021,Whaley2009}.
Another line studies how a chosen estimate is translated into exposure through
volatility-managed portfolios, target-volatility scaling, VIX-based control, or
drawdown-aware rules
\citep{FlemingKirbyOstdiek2001,MoreiraMuir2017,Bozovic2024,NystrupBoydLindstromMadsen2019,MSCI2021,SPDJI2025RC2}.
\method differs by asking a separate question: among several admissible
estimator--controller pairs, which pair should control the portfolio now?

\textbf{Portfolio regimes and execution frictions.}
Dynamic allocation is sensitive to estimation error, regime changes, and trading
costs \citep{Markowitz1952,Hamilton1989,AngBekaert2002,DeMiguelGarlappiUppal2009,AlmgrenChriss2001}.
We keep these concerns explicit through state features, persistence constraints,
and turnover costs, but do not solve a new cross-sectional allocation problem.
Instead, the Router chooses among reusable control policies whose execution rules
are already specified.

\textbf{Routing architectures.}
Mixture-of-experts, sparse conditional computation, contextual bandits, and
model-routing systems all separate context representation from expert selection
\citep{JacobsJordanNowlanHinton1991,LiChuLangfordSchapire2010,ShazeerEtAl2017,FedusZophShazeer2022,OngEtAl2024RouteLLM}.
We transfer this idea to volatility control: the ``experts'' are not neural
subnetworks or language models, but estimator--controller pairs with distinct
risk and execution behavior.

\textbf{LLMs as routers.}
LLM-agent systems show that a language model can inspect structured context,
select tools or experts, and delegate execution
\citep{YaoEtAl2023ReAct,ShenEtAl2023HuggingGPT,WuEtAl2023AutoGen}.
Our main instantiation uses an LLM in this supervisory role. The architecture is
not LLM-specific: the same state, hold/switch, and pair-selection interfaces can
also be implemented by deterministic or learned selectors.

\SectionRef{sec:methodology} turns these connections into the concrete
state--switch--selection interface used by \method.

\section{Methodology}
\label{sec:methodology}

The goal is simple: at each decision date, keep the current control policy if it
still fits the market, or route to a better-suited policy if it does not. The
Router never replaces the downstream portfolio logic. It only decides which
predefined estimator--controller pair is active. This separation is useful for
both interpretability and safety: a routing decision can change the policy, but
portfolio actions remain generated by known control rules.

\subsection{Preliminaries}

We consider a discrete-time investment setting indexed by $t=1,\ldots,T$,
following standard volatility-timing backtests
\citep{FlemingKirbyOstdiek2001,MoreiraMuir2017}. At date $t$, the system observes
only information available by that decision point, denoted $\mathcal I_t$, and
constructs a control-relevant numerical profile
\begin{equation}
s_t = \phi(\mathcal I_t),
\label{eq:state_profile}
\end{equation}
which summarizes quantities such as recent volatility, trend, drawdown, and
execution conditions. Thus, $s_t$ is the measured market context; it is not yet
the Router's final state label.

We define a library of $K$ estimator--controller pairs,
\begin{equation}
\mathcal P=\{p_1,\ldots,p_K\},
\qquad
p_k=(e_k,c_k).
\label{eq:pair_library}
\end{equation}
The two parts have distinct roles. The estimator $e_k$ produces a risk object
$\widehat{\rho}_t^{(k)}$, while the controller $c_k$ converts that risk object
into a portfolio action:
\begin{equation}
\widehat{\rho}_t^{(k)} = e_k(\mathcal I_t),
\end{equation}
\begin{equation}
\mathbf w_t^{(k)}
=
c_k\!\left(\widehat{\rho}_t^{(k)},s_t,\mathbf w_{t-1}\right).
\label{eq:controller}
\end{equation}
For a single asset, $\widehat{\rho}_t^{(k)}$ can be a scalar volatility estimate
and $\mathbf w_t$ becomes a scalar exposure. For Multi-Asset,
$\widehat{\rho}_t^{(k)}$ can be a covariance matrix and $\mathbf w_t$ is a vector
of portfolio weights. A pair therefore represents a complete ``measure risk,
then act'' recipe. Traditional volatility targeting keeps one recipe fixed
\citep{MSCI2021,SPDJI2025RC2}; \method allows the active recipe to change when
the control requirement changes.

\subsection{Routing Formulation}

We formulate volatility control as persistent state-dependent routing over
$\mathcal P$, analogous to expert selection in mixture-of-experts and
model-routing systems
\citep{JacobsJordanNowlanHinton1991,ShazeerEtAl2017,OngEtAl2024RouteLLM}.
The Router separates three questions: \emph{what state are we in?}, \emph{should
we keep the current pair?}, and, only if not, \emph{which pair should replace
it?}

To make this distinction explicit, let $\mathcal I_t^{\mathrm{state}}$ contain
state-only information derived from completed history, and let
$\mathcal I_t^{\mathrm{route}}$ contain routing diagnostics such as trailing
candidate performance, drawdown, tracking error, turnover, and feasibility.
The Router maps the measured profile into a compact state interpretation $z_t$,
reviews whether the active pair $k_{t-1}$ should be held, and selects from the
feasible candidate set $\mathcal C_t\subseteq\mathcal P$ only after a switch:
\begin{align}
z_t
&=\mathcal R_\theta(s_t,\mathcal I_t^{\mathrm{state}}),
\label{eq:router_state}\\
g_t
&=\mathcal G_\theta
\!\left(z_t,\mathcal I_t^{\mathrm{route}},k_{t-1}\right)
\in\{\mathrm{hold},\mathrm{switch}\},
\label{eq:router_gate}\\
k_t
&=
\begin{cases}
k_{t-1}, & g_t=\mathrm{hold},\\
\mathcal S_\theta
\!\left(z_t,\mathcal I_t^{\mathrm{route}},
\mathcal C_t\setminus\{k_{t-1}\}\right),
& g_t=\mathrm{switch}.
\end{cases}
\label{eq:router_select}
\end{align}
Here, $s_t$ is the numerical state profile, while $z_t$ is the Router's compact
interpretation of that profile. The distinction matters because the same raw
measurements can be summarized differently by different Router implementations.

The hold/switch gate is deliberately placed before pair selection. Without this
gate, small changes in candidate rankings could force a new policy every review
step. Persistence constraints such as a minimum holding period or hysteresis can
further prevent unnecessary churn. Once $k_t$ is fixed, routing stops and the
selected pair executes normally:
\begin{equation}
\widehat{\rho}_t=e_{k_t}(\mathcal I_t),
\qquad
\mathbf w_t=c_{k_t}\!\left(\widehat{\rho}_t,s_t,\mathbf w_{t-1}\right).
\label{eq:selected_policy}
\end{equation}
The full decision chain is therefore
\begin{equation}
\mathcal I_t\rightarrow s_t\rightarrow z_t\rightarrow g_t\rightarrow k_t
\rightarrow\mathbf w_t.
\label{eq:routing_chain}
\end{equation}
In words: observe history, summarize the market, interpret the state, decide
whether to switch, select a pair if necessary, and then execute that pair.

Reported portfolio returns charge transaction costs generated by changes in
portfolio weights,
\begin{equation}
R_{t+1}^{\mathrm{net}}
=
\mathbf w_t^\top \mathbf r_{t+1}
-
\lambda_{\mathrm{to}}\lVert\mathbf w_t-\mathbf w_{t-1}\rVert_1,
\label{eq:net_return}
\end{equation}
with the corresponding cash leg included in Multi-Asset. Pair switching is
regularized by the hold/switch gate and persistence rules rather than by a
uniform extra fee for changing pair identity. \AppendixRef{app:switching}
provides the implementation-level accounting.

All routing diagnostics are lagged. For each candidate, the Router uses trailing
history
\begin{equation}
\mathcal H_t=\{t-H,\ldots,t-1\},\qquad H=63,
\end{equation}
so the decision at $t$ never contains $r_{t+1}$ or any later portfolio outcome.
\AppendixRef{app:noleak} gives the exact timing convention.

\subsection{LLM-Based Routing}

The same interface can be implemented by rules, learned selectors, contextual
bandits, or an LLM. In our LLM instantiation, the model acts as a constrained
routing agent. It does not see an unconstrained natural-language trading task.
Instead, it receives structured records built from decision-time information and
performs the same three steps defined above.

First, it returns a valid state label. Second, it returns exactly
\texttt{hold} or \texttt{switch}. Third, only after \texttt{switch}, it returns
one pair identifier from the supplied candidate set. It does not predict returns,
change controller parameters, or directly output portfolio weights. All responses
are parsed as JSON and checked against closed admissible sets; invalid or failed
responses use deterministic fallback logic. The prompt schemas, parsers, and
fallbacks are in \AppendixRef{app:prompt_routing_format}, and the full routing
specification is in \AppendixRef{app:method}.

\begin{figure*}[t]
    \centering
    \includegraphics[width=1\textwidth]{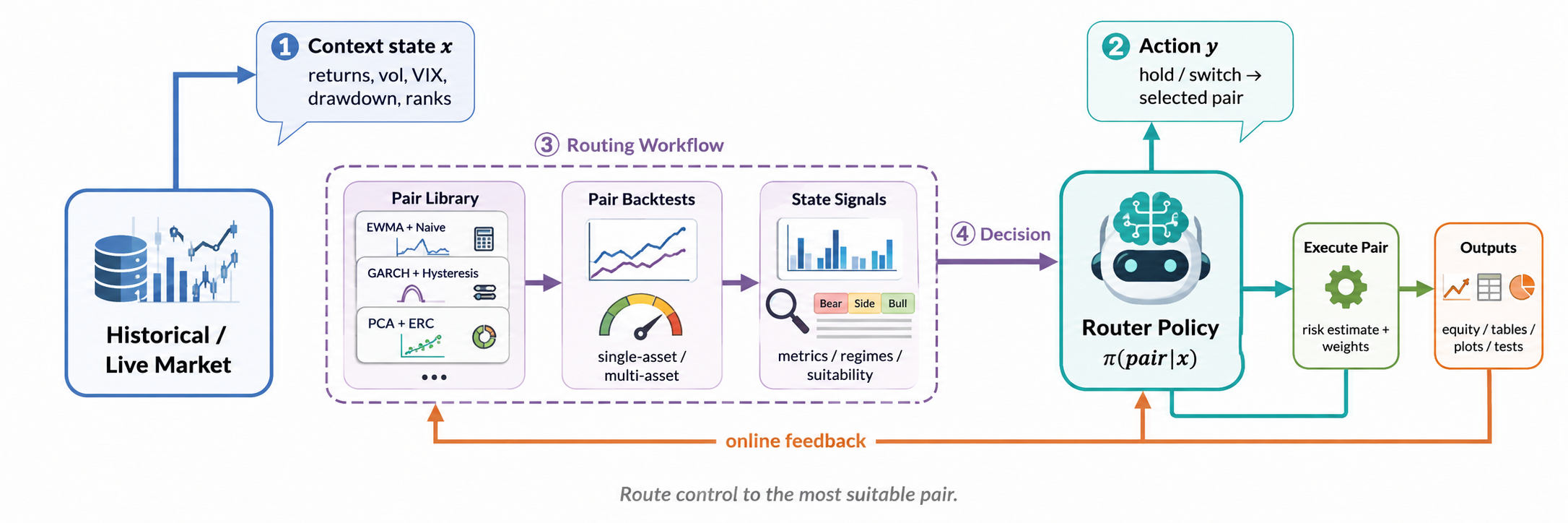}
    \Description{VolRouter pipeline from market information to a state profile, switch review and pair selection, selected estimator and controller, and final portfolio action.}
    \caption{
    Overview of \method. Market history is summarized into a control-relevant
    profile. The Router interprets the state, decides whether the active pair
    should be kept, and selects a replacement only when switching is justified.
    The selected estimator--controller pair then produces the portfolio action;
    the LLM, when used, never produces the trade directly.
    }
    \label{fig:volrouter_framework}
\end{figure*}

\SectionRef{sec:experimental_setup} evaluates whether this explicit routing layer
improves risk-adjusted control, which parts of the three-stage decision matter,
and how sensitive the result is to deployment choices and Router backbones.


\section{Experimental Setup}
\label{sec:experimental_setup}

The experiments answer four questions in order: does routing improve
risk-adjusted control over fixed and adaptive alternatives; which parts of the
routing decision matter; how sensitive is performance to switching, execution
cost, risk target, and library size; and does the behavior persist across Router
backbones? This order mirrors the framework in \SectionRef{sec:methodology}: we
first test the complete system, then remove components, vary deployment choices,
and finally change the routing model.

\subsection{Tasks and Datasets}

We evaluate \method in four out-of-sample settings chosen to expose different
control demands. \textbf{S\&P 500} is a single-asset volatility-targeting task on
broad U.S. equity exposure. \textbf{Multi-Asset} routes over covariance-estimator
and portfolio-controller pairs in a diversified cross-asset portfolio.
\textbf{Bitcoin} is a high-volatility stress test with large risk shifts, while
\textbf{USDT} is a low-volatility boundary case where aggressive adaptation may
be unnecessary. The goal is not to compare asset classes by their absolute
Sharpe ratios, but to ask whether routing helps \emph{within} each control
environment.

The settings use different risk budgets because their natural volatility scales
are different. S\&P 500 uses a 10\% volatility target, 5 bp transaction cost,
252-day estimation window, and 252-day annualization; Multi-Asset uses 10\%,
5 bp, 126 days, and 252 days; Bitcoin uses 35\%, 8 bp, 90 days, and 365 days;
and USDT uses 2\%, 2 bp, 90 days, and 365 days, respectively. Estimator inputs,
state features, candidate construction, and calibration use only information
available before the evaluated action. The action formed at date $t$ is applied
to subsequent returns. Preprocessing and timing details are given in
\AppendixRef{app:data}, \AppendixRef{app:protocol}, and
\AppendixRef{app:noleak}.

\subsection{Baselines and Evaluation}

Routing methods draw from the same eligible estimator--controller library; fixed
baselines use their corresponding fixed policies. The library combines
realized/EWMA, GARCH/HAR, machine-learning, and market-specific risk estimators
with scaling, regime-aware, trend, drawdown, tail-risk, and multi-asset
controllers. \AppendixRef{app:library} lists the implementation-level library and
eligibility rules.

The baseline progression is intentional. \textbf{RV + Naive Scaling} and
\textbf{EWMA Targeting} ask what happens when one standard pair is used
throughout. \textbf{Regime-Aware Fixed} asks whether state dependence inside one
predefined controller is already sufficient. \textbf{Mixture-of-Experts} softly
combines specialized policies, whereas \textbf{Contextual Bandit} explicitly
selects a strategy from context. \method differs by separating the decision to
\emph{switch} from the decision of \emph{which replacement to choose}, and by
maintaining the selected pair persistently between review decisions. Where
applicable, adaptive baselines receive the same state information, eligible
library, historical diagnostics, and pre-evaluation calibration data.

We report annualized return, annualized volatility, Sharpe ratio, maximum
drawdown (MDD), and daily 95\% CVaR. These metrics must be read jointly: lower
volatility is not automatically better because each setting has an explicit risk
target, and a very low-volatility strategy can simply be under-exposed.
Secondary analyses use Sortino ratio, rolling Sharpe, turnover, volatility
tracking, switching behavior, and sensitivity diagnostics. All portfolio returns
include the setting-specific transaction costs above; metric definitions are in
\AppendixRef{app:metrics}.


\section{Results}
\label{sec:results}


\subsection{Overall Performance}
\label{sec:overall_results}

\begin{table*}[t!]
\centering
\caption{
\textbf{Main out-of-sample results across four volatility-control
settings.}
Ann. Ret. and Ann. Vol. denote annualized return and annualized
volatility; MDD denotes maximum drawdown; and CVaR denotes daily
95\% conditional value-at-risk.
Higher annualized return and Sharpe are better, whereas lower MDD and
CVaR are better.
Annualized volatility should be interpreted relative to the
setting-specific volatility target rather than minimized
unconditionally.
Within each market block, colors are measured relative to
RV + Naive Scaling.
}
\label{tab:main_results}
\footnotesize
\setlength{\tabcolsep}{7.0pt}
\renewcommand{\arraystretch}{1.12}

\begin{tabular*}{\textwidth}{@{\extracolsep{\fill}}l ccccc@{}}
\toprule
\tableheader{Method}
& \tableheader{Ann. Ret. (\%)}
& \tableheader{Ann. Vol. (\%)}
& \tableheader{Sharpe}
& \tableheader{MDD (\%)}
& \tableheader{CVaR (\%)} \\
\midrule

\blockrow{S\&P 500} \\
RV + Naive Scaling
& 11.11 & 11.67 & 0.952 & 15.10 & 1.76 \\
EWMA Targeting
& \lossone{10.31} & \gainthree{10.85} & \lossone{0.951}
& \gainone{14.02} & \gainone{1.63} \\
Regime-Aware Fixed
& \losstwo{10.28} & \losstwo{12.39} & \losstwo{0.830}
& \lossone{15.85} & \lossone{1.81} \\
Mixture-of-Experts
& \lossone{10.45} & \gainthree{10.90} & \gainone{0.958}
& \gainone{15.06} & \gainone{1.66} \\
Contextual Bandit
& \losstwo{9.35} & \gainthree{10.45} & \lossone{0.895}
& \gainone{13.41} & \gainone{1.60} \\
\methodlabel
& \lossone{\textbf{10.84}}
& \gainthree{\textbf{8.87}}
& \gainthree{\textbf{1.222}}
& \gaintwo{\textbf{12.58}}
& \gainthree{\textbf{1.32}} \\

\addlinespace[4pt]
\blockrow{Multi-Asset} \\
RV + Naive Scaling
& 16.35 & 10.91 & 1.498 & 7.84 & 1.56 \\
EWMA Targeting
& \lossthree{11.62} & \losstwo{9.05} & \losstwo{1.283}
& \lossone{8.88} & \gainone{1.32} \\
Regime-Aware Fixed
& \lossthree{11.60} & \losstwo{8.86} & \losstwo{1.309}
& \gainone{6.26} & \gainone{1.30} \\
Mixture-of-Experts
& \losstwo{12.41} & \losstwo{8.88} & \lossone{1.398}
& \gainone{6.28} & \gainone{1.27} \\
Contextual Bandit
& \lossthree{10.59} & \lossthree{8.04} & \losstwo{1.318}
& \lossone{9.24} & \gaintwo{1.20} \\
\methodlabel
& \losstwo{\textbf{13.20}}
& \losstwo{\textbf{8.57}}
& \gainone{\textbf{1.540}}
& \gainone{\textbf{6.60}}
& \gaintwo{\textbf{1.18}} \\

\addlinespace[4pt]
\blockrow{Bitcoin} \\
RV + Naive Scaling
& 29.56 & 39.57 & 0.747 & 51.69 & 4.68 \\
EWMA Targeting
& \lossone{27.26} & \gainthree{36.56} & \lossone{0.745}
& \lossone{51.95} & \gainone{4.30} \\
Regime-Aware Fixed
& \lossthree{1.60} & \lossthree{6.73} & \lossthree{0.238}
& \gainthree{7.24} & \gainthree{0.99} \\
Mixture-of-Experts
& \lossthree{5.99} & \lossthree{7.51} & \gainone{0.798}
& \gainthree{6.85} & \gainthree{1.00} \\
Contextual Bandit
& \lossthree{10.28} & \gainthree{36.27} & \lossthree{0.283}
& \lossone{51.47} & \gainone{4.56} \\
\methodlabel
& \gainone{\textbf{30.81}}
& \lossthree{\textbf{27.43}}
& \gainthree{\textbf{1.123}}
& \gaintwo{\textbf{39.39}}
& \gaintwo{\textbf{3.25}} \\

\addlinespace[4pt]
\blockrow{USDT} \\
RV + Naive Scaling
& -4.71 & 4.37 & -1.078 & 17.75 & 0.32 \\
EWMA Targeting
& \gainone{-3.81} & \gaintwo{2.99} & \losstwo{-1.274}
& \gainone{15.95} & \gainone{0.27} \\
Regime-Aware Fixed
& \gainthree{4.75} & \lossthree{0.52} & \gainthree{9.140}
& \gainthree{0.89} & \gainthree{0.05} \\
Mixture-of-Experts
& \gainthree{4.10} & \gainthree{1.04} & \gainthree{3.943}
& \gainthree{2.70} & \gainthree{0.08} \\
Contextual Bandit
& \gainthree{4.94} & \lossthree{0.52} & \gainthree{9.566}
& \gainthree{0.81} & \gainthree{0.05} \\
\methodlabel
& \gainthree{\textbf{4.41}}
& \lossthree{\textbf{0.53}}
& \gainthree{\textbf{8.379}}
& \gainthree{\textbf{1.25}}
& \gainthree{\textbf{0.06}} \\

\bottomrule
\end{tabular*}
\end{table*}

Table~\ref{tab:main_results} shows a consistent pattern in three settings:
routing improves risk-adjusted performance mainly by changing risk deployment,
not by mechanically raising raw return. On S\&P 500, Sharpe rises from 0.952 for
RV + Naive Scaling to 1.222 while MDD falls from 15.10\% to 12.58\% and CVaR from
1.76\% to 1.32\%. Annualized return is slightly lower (10.84\% versus 11.11\%),
so the improvement is not explained by higher exposure.

Multi-Asset shows the same trade-off. RV + Naive Scaling has the highest raw
return (16.35\%), but \method reaches a higher Sharpe (1.540 versus 1.498) and a
lower CVaR (1.18\% versus 1.56\%). In other words, the Router gives up some raw
return while using the risk budget more efficiently across market states.

Bitcoin is the strongest case: \method reaches 30.81\% annualized return and a
1.123 Sharpe while reducing MDD from 51.69\% to 39.39\%. Regime-Aware Fixed and
Mixture-of-Experts have smaller drawdowns, but their realized volatilities are
only 6.73\% and 7.51\% against a 35\% target, indicating substantial
under-exposure. \method remains much closer to the intended risk budget.

USDT is the boundary case. Contextual Bandit and Regime-Aware Fixed achieve
higher Sharpe than \method, and all three operate at realized volatility far
below the 2\% target. We therefore read this setting as evidence that an explicit
pair-routing layer is not always necessary: when the environment is unusually
stable, a simpler state-aware selector can already capture much of the benefit.



\begin{figure*}[t]
    \centering
    \includegraphics[width=0.98\textwidth]
    {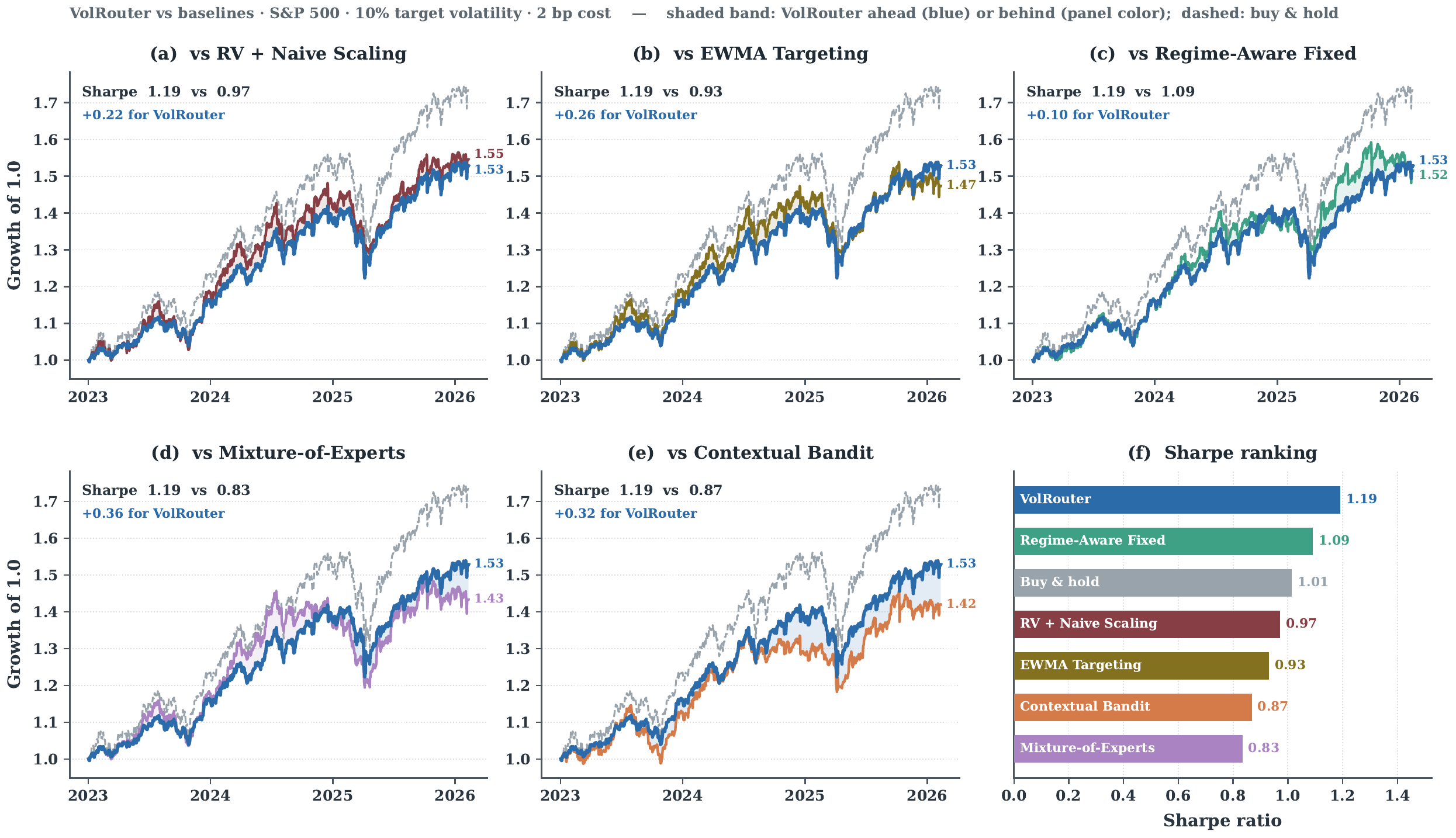}
    \Description{
    Six-panel comparison of VolRouter with S\&P 500 volatility-control
    baselines. Five panels show cumulative wealth paths and the final
    panel ranks methods by Sharpe ratio.
    }
    \caption{
    \textbf{Path-level S\&P 500 diagnostic} under a 10\% volatility target
    and 2 bp costs. Panels (a)--(e) compare cumulative wealth and panel (f)
    reports Sharpe ratios. This run is separate from the 5 bp headline
    setting in Table~\ref{tab:main_results} and is used only to show when
    path-level differences accumulate.
    }
    \label{fig:equity_compare}
\end{figure*}

Figure~\ref{fig:equity_compare} is deliberately a separate 2 bp path-level
diagnostic, not a second estimate of the 5 bp headline table. Within that run,
\method stays close to strong alternatives for long periods and separates in
specific episodes, showing how repeated local routing decisions accumulate into
different portfolio paths.


\subsection{What Makes Routing Effective?}
\label{sec:ablation_results}

\begin{figure*}[t]
    \centering
    \includegraphics[width=0.94\textwidth]
    {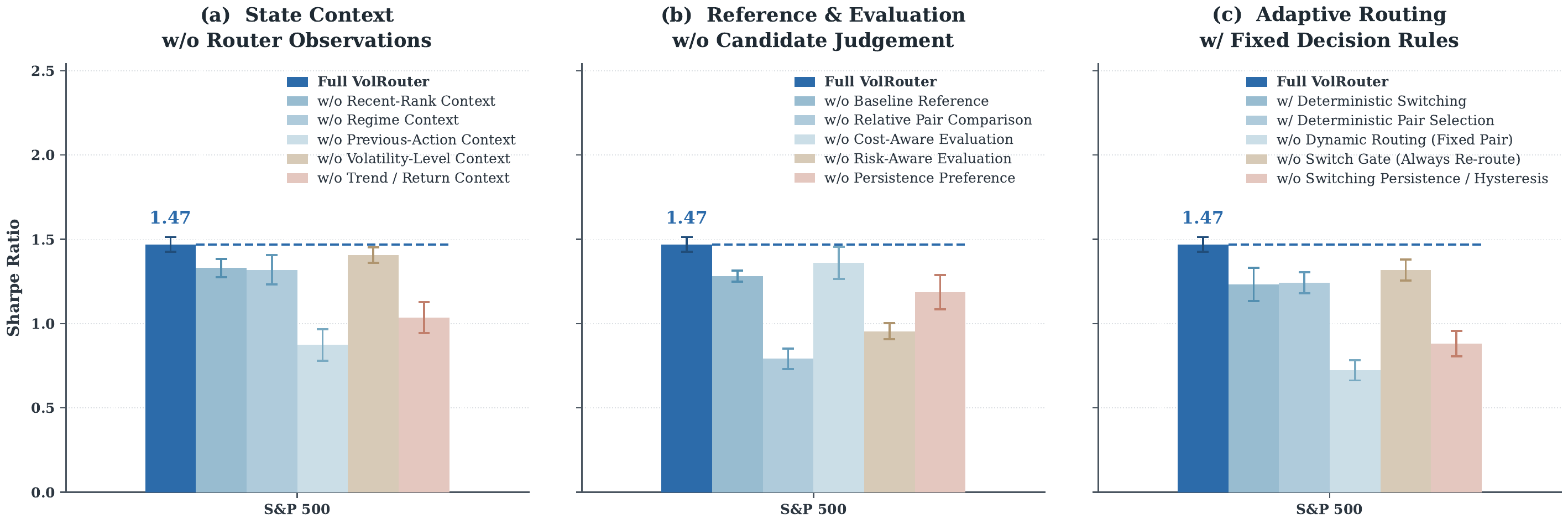}
    \Description{
    Component ablations of VolRouter on S&P 500 grouped into state
    context, reference and candidate evaluation, and adaptive routing.
    }
    \caption{
    \textbf{Component ablation on S\&P 500.}
    Panels remove (a) state context, (b) candidate-reference/evaluation
    signals, and (c) adaptive-routing components. Full \textit{VolRouter}
    has Sharpe 1.47 under this ablation protocol. Error bars summarize the
    ablation driver and are not confidence intervals for Table~\ref{tab:main_results}.
    }
    \label{fig:ablation_sp500_sharpe}
\end{figure*}

The ablation driver has its own evaluation protocol, so its Sharpe values are
compared only within Figure~\ref{fig:ablation_sp500_sharpe}, not directly with
the headline table. The pattern is nevertheless clear. Removing temporal state
context weakens routing because the Router loses information about how the
current policy relates to the recent market path. Removing relative candidate
comparison or risk-aware evaluation also degrades performance, showing that
policies must be judged against one another under the current state rather than
by an isolated unconditional score.

The adaptive-routing ablations are especially informative. Replacing dynamic
routing with a fixed pair causes one of the largest losses, while removing
persistence or replacing adaptive decisions with deterministic alternatives also
reduces Sharpe. The gain therefore comes from the combination of state context,
relative candidate evaluation, and persistent but revisable policy selection---
not simply from exposing the system to a larger library.


\subsection{Sensitivity and Robustness}
\label{sec:sensitivity_results}

\begin{figure*}[t]
    \centering
    \includegraphics[width=0.92\textwidth]
    {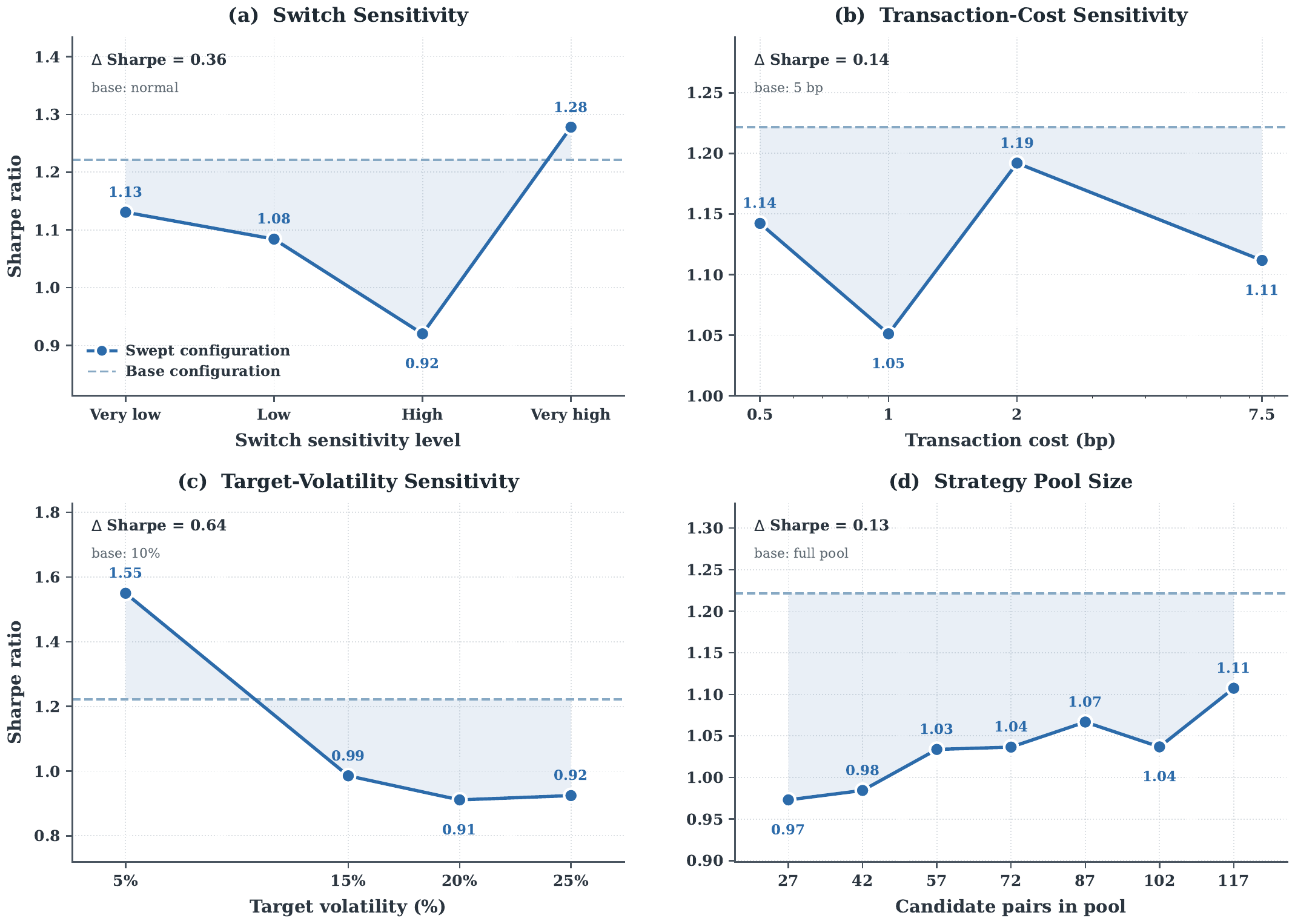}
    \Description{
    Four sensitivity analyses varying switching aggressiveness,
    transaction cost, target volatility, and estimator-controller
    candidate-pool size.
    }
    \caption{
    \textbf{Sensitivity of \textit{VolRouter}.}
    We vary (a) switching sensitivity, (b) transaction cost, (c) target
    volatility, and (d) candidate-pool size one at a time. Dashed lines
    mark the canonical base configuration; the base need not be one of the
    displayed sweep points.
    }
    \label{fig:router_sensitivity}
\end{figure*}

Figure~\ref{fig:router_sensitivity} varies one deployment choice at a time;
the four panels are separate diagnostics rather than one shared numerical grid.
Switching sensitivity changes how readily the Router reacts, transaction cost
changes execution friction, target volatility changes the risk budget, and
candidate-pool size changes the available policy diversity.

Target volatility produces the largest movement: Sharpe is 1.55 at the 5\%
target and falls toward roughly 0.9--1.0 at the more aggressive 15--25\%
targets. Transaction-cost changes are narrower, while switching sensitivity is
non-monotonic. Candidate-pool expansion helps overall but not at every
intermediate size. These results have a common interpretation: more risk, more
switching, or more candidate policies do not automatically improve control; the
Router must use that additional freedom selectively.


\subsection{Router Backbones and Cross-Market Stability}
\label{sec:backbone_results}

\begin{figure*}[t]
    \centering
    \includegraphics[width=0.96\textwidth]
    {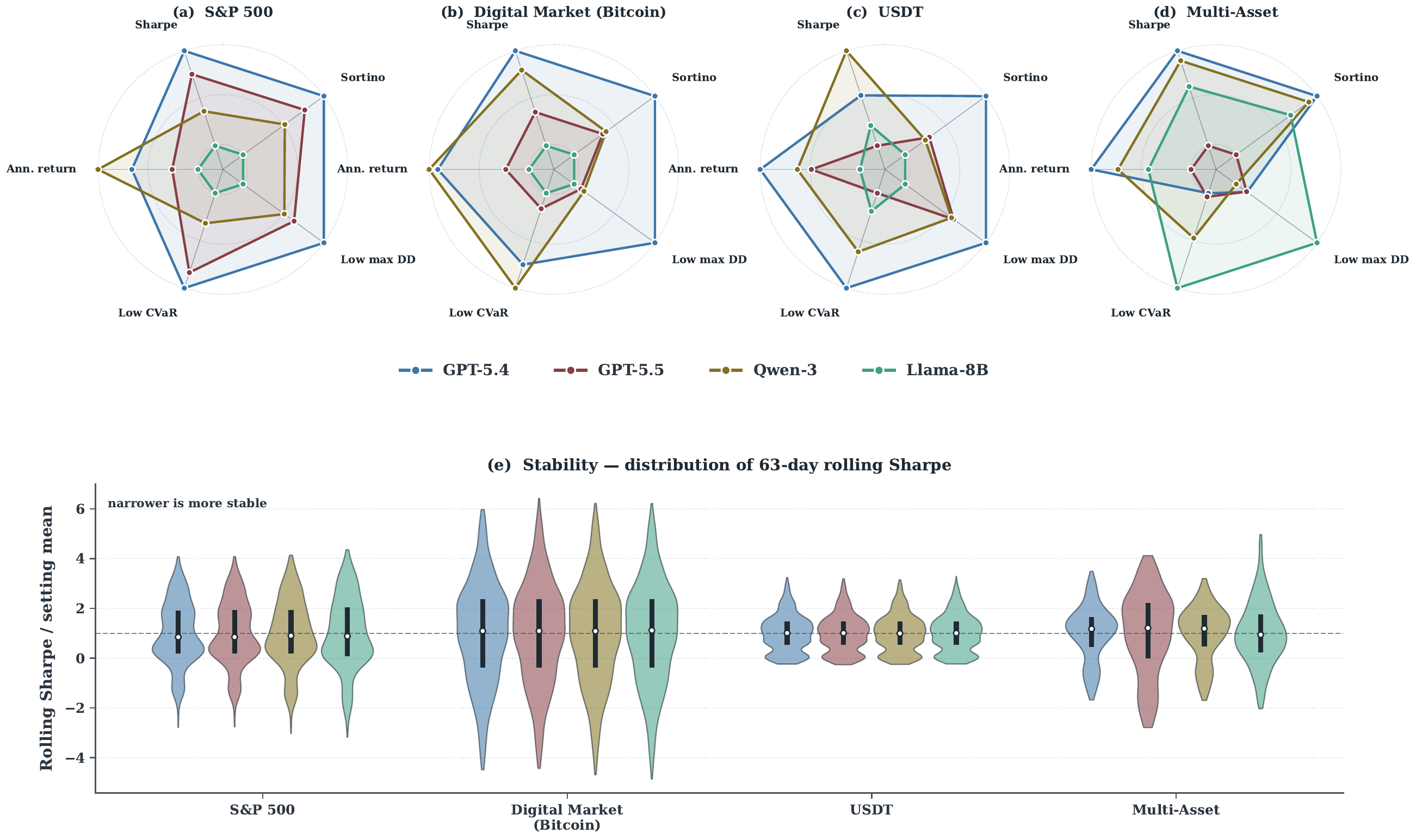}
    \Description{
    Radar plots comparing four router language-model backbones across
    four markets and a violin-style distribution of 63-day rolling
    Sharpe ratios.
    }
    \caption{
    \textbf{Router backbones across markets.}
    Panels (a)--(d) compare normalized return and risk metrics; panel (e)
    shows 63-day rolling-Sharpe distributions after within-setting
    normalization. In the S\&P 500 sweep, the regime series is held fixed,
    isolating downstream switch and pair-selection behavior.
    }
    \label{fig:router_models}
\end{figure*}

Figure~\ref{fig:router_models} asks whether routing behavior depends on one
language-model backbone. In the S\&P 500 sweep, the regime series is held fixed,
so that panel isolates switch review and pair selection rather than end-to-end
state inference. Backbone rankings change across markets and risk dimensions;
no model is uniformly strongest. Rolling-Sharpe distributions also vary by
market, with Bitcoin showing the widest temporal variation and USDT the
narrowest. Useful routing behavior nevertheless appears across multiple
backbones, indicating that the reported effect is not tied to one specific LLM.

\section{Analysis}
\label{sec:analysis}

The results support a more useful conclusion than ``routing improves Sharpe.''
They show \emph{what the Router is actually buying}: better timing of control
policies, not a new source of return prediction.

\textbf{Insight 1: the gain is mainly state-dependent risk deployment, not
alpha forecasting.}
On S\&P 500, \method raises Sharpe from 0.952 to 1.222 even though annualized
return is slightly lower; the improvement comes with lower realized volatility,
drawdown, and CVaR. Multi-Asset shows the same pattern: raw return is below RV +
Naive Scaling, but Sharpe is higher and CVaR is lower. Bitcoin provides the
complementary case, where return rises while realized volatility and drawdown
fall. Across these settings, the common mechanism is therefore not ``predict a
higher return.'' The Router changes \emph{which validated risk response is active
when the cost of being too aggressive or too defensive changes}. This also
clarifies why an LLM can be useful without forecasting the next return: it is
choosing among known control recipes rather than creating an alpha signal.

\textbf{Insight 2: market-state classification alone is insufficient; routing
is an action-conditioned comparison problem.}
Figure~\ref{fig:ablation_sp500_sharpe} shows that removing previous-action
context is more damaging than removing several individual state descriptors,
and removing relative pair comparison causes one of the largest losses in the
reference/evaluation block. The relevant question is therefore not only ``what
state are we in?'' but ``given this state, is the policy already in control still
the right one relative to the alternatives?'' A high-volatility label, for
example, does not imply one universal action if the active controller is already
handling the risk well. This result is the main reason to keep state inference,
switch review, and pair selection separate: directly mapping each state label to
one fixed policy discards information about the policy currently being executed.

\textbf{Insight 3: persistence is a form of temporal regularization, not merely
a way to save transaction costs.}
The adaptive-routing ablations and the switching sweep tell the same story.
Removing persistence/hysteresis reduces Sharpe, and switching sensitivity is
non-monotonic rather than ``more reactive is better.'' At the same time, the
transaction-cost sweep in Figure~\ref{fig:router_sensitivity} moves Sharpe less
than the switching-sensitivity sweep. Persistence is therefore doing more than
avoiding fees: it filters short-lived changes in candidate rankings and prevents
policy churn caused by noisy local evidence. A router that simply takes the
current argmax at every review step is not equivalent to \method, even if both
see the same candidate scores. The hold/switch gate is useful because it asks
for evidence that the existing policy has become inappropriate before reopening
the selection problem.

\textbf{Insight 4: the risk budget is first-order; Router tuning is second-order.}
Among the four deployment sweeps, target volatility produces the largest Sharpe
range (about 0.64), compared with about 0.36 for switching sensitivity and only
about 0.14 and 0.13 for transaction cost and candidate-pool size. This gives a
clear deployment hierarchy. The risk target determines how much risk the system
is allowed to use; the policy library determines which responses are available;
the Router only decides how to move among those responses. A sophisticated
Router therefore cannot rescue an inappropriate risk budget. In practice, the
risk target should be selected and validated before tuning switching behavior or
expanding the candidate pool.

\textbf{Insight 5: useful diversity matters more than a large library.}
The candidate-pool curve improves overall but is not monotonic. Adding policies
can help, yet some intermediate expansions add little or temporarily reduce
Sharpe. This rules out a simple ``more experts is better'' explanation. A useful
candidate must contribute a behavior that is both different and identifiable
from lagged evidence---for example, a different reaction speed, drawdown brake,
tail-risk response, or covariance treatment. Redundant or noisy pairs enlarge
the choice set without adding a new control capability. The design objective is
therefore \emph{behavioral contrast}, not library cardinality.

\textbf{Insight 6: the routing premium is conditional, and USDT shows the
boundary.}
USDT is informative because \method is not the best method there. Regime-Aware
Fixed and Contextual Bandit obtain higher Sharpe, while all strong methods
realize volatility far below the 2\% target. In a stable environment, candidate
policies can disagree less and a simpler selector can capture most of the useful
state dependence. The case also exposes an evaluation pitfall: Sharpe can reward
a method that simply stays far below the intended risk budget. Risk-adjusted
performance should therefore be read jointly with risk-target adherence rather
than in isolation.

Finally, Figure~\ref{fig:router_models} shows that no Router backbone dominates
across all markets and risk dimensions. The transferable object is therefore
better viewed as the \emph{routing interface}---structured state, an explicit
hold/switch decision, and closed-set pair selection---rather than one universally
best language model. Backbone choice remains setting-dependent, but it can change
without redefining the downstream estimators or controllers.

Overall, routing is most promising when the risk budget is sensible, candidate
policies make genuinely different state-dependent trade-offs, and switching is
persistent enough to ignore short-lived ranking noise. The analysis is
descriptive of held-out point estimates; statistical scope is given in
\AppendixRef{app:reliability}.

\FloatBarrier

\section{Conclusion}
\label{sec:conclusion}

We presented \method, a persistent Router over validated
estimator--controller pairs. Its value comes from selecting \emph{when}
different risk-control rules should act, not from predicting returns directly.
Routing helps most when candidate policies make meaningful state-dependent
trade-offs; it helps less when markets are stable or the library adds little
behavioral contrast. Volatility management therefore concerns both estimating
risk and choosing which validated response should act now.



\appendix

\newcommand{\promptgap}{\par\noindent\mbox{}\par}
\newenvironment{routerpromptbox}[1]
{%
  \par\noindent\begin{minipage}{\linewidth}
  \hrule
  \rule{0pt}{2.0ex}\textbf{#1}\par
  \begingroup\small\ttfamily\raggedright\sloppy\setlength{\parindent}{0pt}%
}
{%
  \par\endgroup\rule{0pt}{1.2ex}\hrule
  \end{minipage}\par
}

\section{Full Routing Specification}
\label{app:method}

\subsection{Unified Routing Formulation}

The appendix uses the same objects as the main methodology. Let
\[
\mathcal{P}=\{p_k\}_{k=1}^{K},\qquad p_k=(E_k,C_k),
\]
be the candidate library of estimator--controller pairs. For a single-asset
task, $E_k$ maps lagged market data to a scalar volatility estimate; for a
multi-asset task, it maps lagged data to a covariance estimate. The controller
$C_k$ then maps that risk object, the target risk budget, and the previous
portfolio state to an executable action.

At date $t$, the observable history $\mathcal I_t$ is first converted into a
numerical state profile $s_t=\phi(\mathcal I_t)$. We distinguish this measured
profile from the Router's compact interpretation $z_t$. The state input
$\mathcal I_t^{\mathrm{state}}$ contains only lagged market information used to
interpret the environment. The routing input $\mathcal I_t^{\mathrm{route}}$
adds lagged diagnostics for the active and candidate pairs. This separation is
useful because market-state inference should not be defined by future outcomes or
by an unconstrained look at candidate performance.

Both implementations use the same three-stage decision:
\[
\text{state inference}\;\rightarrow\;\text{switch review}\;\rightarrow\;\text{pair selection}.
\]
Let $k_{t-1}$ be the active pair and $\mathcal C_t\subseteq\mathcal P$ the
eligible candidate set. We write
\begin{align}
z_t
&=\mathcal{R}_{\theta}(s_t,\mathcal{I}^{\mathrm{state}}_t),
\label{eq:app_regime}\\
g_t
&=\mathcal{G}_{\theta}
\left(z_t,\mathcal{I}^{\mathrm{route}}_t,k_{t-1}\right)
\in\{\mathrm{hold},\mathrm{switch}\},
\label{eq:app_switch}\\
k_t
&=
\begin{cases}
k_{t-1}, & g_t=\mathrm{hold},\\[2mm]
\mathcal{S}_{\theta}
\left(z_t,\mathcal{I}^{\mathrm{route}}_t,
\mathcal{C}_t\setminus\{k_{t-1}\}\right),
& g_t=\mathrm{switch}.
\end{cases}
\label{eq:app_select}
\end{align}
For the single-asset Router,
\[
z_t\in\{\mathrm{low},\mathrm{middle},\mathrm{high}\},
\]
whereas the Multi-Asset Router uses
\[
z_t\in\{\mathrm{risk\_on},\mathrm{balanced},\mathrm{defensive}\}.
\]
A new pair is therefore never chosen just because another candidate has a
slightly higher recent score: pair selection is called only after the explicit
hold/switch decision.

A persistence constraint is applied through a sticky period. A switch is
admissible only when
\[
a_t+1\ge H_{\mathrm{sticky}},
\]
where $a_t$ is the age of the active pair. The Multi-Asset Router additionally
maintains cooldown and hold-gate state. These mechanisms implement the same idea
as the main text: policy changes should be possible, but not automatic at every
review step.

\subsection{Execution and Deterministic Reference}

The equations above determine \emph{which} pair is active. They do not redefine
the downstream estimators or controllers. Their exact equations, portfolio
execution rules, transaction-cost accounting, and deterministic fallback score
are collected in \AppendixRef{app:mathematical_specification}. Keeping these
implementation details in one place prevents the method overview and the library
specification from silently using different conventions.

\section{Policy Library}
\label{app:library}

\subsection{Single-Asset Estimators}

The repository contains autoregressive, realized-volatility, EWMA,
GARCH/GJR-GARCH, HAR, regime-aware, machine-learning, ensemble, and
crypto-specific estimators.  The effective families used by the router
include AR(1), AR(2), EWMA, RealizedVol, NaiveVolEstimator, GARCH,
GJR-GARCH, Regime GJR-GARCH, HAR-RV, HAR-RV with rates, regime HAR,
Hybrid EWMA Regime, LightGBM, Random Forest, RNN/MLP, XGB-VIX,
Dynamic Precision Ensemble, Intraday Realized Volatility,
Range-Based Volatility, and Crypto Composite Volatility.

Important defaults include EWMA half-life 20; HAR weekly/monthly horizons
5/22; machine-learning refit intervals of 21 trading days in several
estimators; annualization 252 for the equity setting and 365 for the
crypto estimators.  The wavelet implementation is commented out and is
not an effective candidate.  The master configuration also excludes
Lasso/wavelet and buy-and-hold estimator identifiers from router
eligibility.

\subsection{Single-Asset Controllers}

The controller library includes constant weight, naive scaling, clipped
volatility targeting, hysteresis control, trend filtering, variance scaling,
regime-switch control, drawdown braking, drawdown modulation, CVaR/ES
targeting, priority-stack control, shock throttling, and peg-aware control.
The naive controller implements the familiar inverse-volatility rule
\begin{equation}
w_t^{\mathrm{naive}}
=\frac{\sigma^\star}{\widehat\sigma_t},
\end{equation}
subject to global exposure bounds and a hard-coded no-trade band of
$0.05$.

\subsection{Multi-Asset Estimators and Controllers}

The multi-asset covariance library includes sample, expanding, EWMA, and
diagonal EWMA estimators; rolling-correlation and shrinkage estimators; and
Ledoit--Wolf, downside, robust-median, regime-switching, VIX-scaled, PCA, and
dynamic-blend variants. Representative defaults include EWMA half-life 21,
shrinkage 0.25, and three PCA components.

Portfolio controllers include equal weight, buy-and-hold, inverse
volatility, minimum variance, volatility-capped minimum variance, equal
risk contribution, diversified risk parity, momentum tilt, mean
variance, regime-aware risk budgeting, drawdown brake, and hysteresis
portfolio control.

The full single-asset master configuration contains 209 named
estimator--controller pairs before experiment-specific eligibility filters.
Only executable and eligible pairs enter a given routing candidate set;
non-functional or explicitly excluded implementations, including the
commented-out wavelet path, are not treated as active candidates.

\section{Data and Preprocessing}
\label{app:data}

\begin{table*}[t]
\centering
\caption{Canonical evaluation settings and backtest constants.}
\label{tab:app_settings}
\scriptsize
\setlength{\tabcolsep}{4pt}
\begin{tabular}{llcccc}
\toprule
\textbf{Setting} & \textbf{Instrument / Universe}
& $\sigma^\star$ & \textbf{Cost (bps)}
& \textbf{Annualization} & \textbf{Rolling Window} \\
\midrule
S\&P 500 & SPY / ES futures & 0.10 & 5 & 252 & 252 \\
Multi-Asset & cross-asset panel & 0.10 & 5 & 252 & 126 \\
Digital Market (Bitcoin) & BTCUSD & 0.35 & 8 & 365 & 90 \\
USDT & USDTUSD & 0.02 & 2 & 365 & 90 \\
\bottomrule
\end{tabular}
\end{table*}

\subsection{S\&P 500 / ES Pipeline}

The equity pipeline parses and sorts timestamps, removes duplicate rows
and non-positive prices, filters ES contracts, resamples minute data to
business-day frequency, detects futures roll dates, constructs a
back-adjusted continuous close series, reindexes to a business-day
calendar, and forward-fills the continuous price and contract identifier.
Raw log returns are
\begin{equation}
r_t=\log\frac{P_t^{\mathrm{adj}}}{P_{t-1}^{\mathrm{adj}}}.
\end{equation}
A 252-observation rolling standard deviation (minimum 20 observations),
denoted $\sigma^{\mathrm{roll}}_t$, is used for dynamic winsorization:
\begin{equation}
r_t^c
=\operatorname{clip}
\left(r_t,-5\sigma^{\mathrm{roll}}_t,+5\sigma^{\mathrm{roll}}_t\right).
\end{equation}
This local rolling-volatility symbol is distinct from the Router state profile
$s_t$ used in the main methodology.
The cleaned series $r_t^c$ is used by volatility estimators and market
features, while raw $r_t$ is retained for P\&L.  This distinction is
intentional and should be preserved in reproduction.

\subsection{Multi-Asset Panel}

The processed panel is read as numeric returns, rows that are entirely
missing are removed, and remaining missing entries are set to zero:
\begin{equation}
r_{t,i}\leftarrow
\begin{cases}
r_{t,i}, & \text{observed},\\
0, & \text{missing}.
\end{cases}
\end{equation}
This is equivalent to treating a missing asset observation as a flat
return for that day. This implementation choice applies before covariance
estimation and should be considered when interpreting Multi-Asset results.

\subsection{Risk-Free Rate}

The multi-asset engine reads the FRED three-month Treasury yield and
converts it to a daily log return:
\begin{equation}
r_t^f
=\log\left(1+\frac{\mathrm{DGS3MO}_t/100}{252}\right).
\end{equation}
The series is forward-filled, with any remaining missing values replaced
by zero.

\section{Mathematical Specification of Estimators and Controllers}
\label{app:mathematical_specification}

This section gives the implementation-level mathematical specification of the
policy library.  Equations follow the audited code paths rather than idealized
textbook definitions.  Generic time indices such as $s$ describe an estimator's
internal recursion; any forecast denoted by $\widehat\sigma_t$ or
$\widehat\Sigma_t$ is formed only from information available strictly before
the routing decision at $t$ unless an explicit audit caveat is stated.

\subsection{Notation and Timing Convention}

Let $r_t$ denote the log return on day $t$, $W$ the estimation window,
$A$ the annualization factor, $\widehat\sigma_t$ the annualized volatility
forecast, $\widehat\Sigma_t$ the annualized covariance forecast, $v^\star$ the
annualized target volatility, and $w_t$ the portfolio exposure formed at $t$.
For multi-asset strategies, $\mathbf w_t$ denotes the risky-asset weight vector.
The universal timing convention is
\begin{align}
\widehat\sigma_t,\widehat\Sigma_t
&:\ \text{data strictly before }t,\\
w_t,\mathbf w_t
&:\ \text{applied to returns at }t+1.
\label{eq:timing_convention}
\end{align}
The annualization factor is $A=252$ for the equity and multi-asset settings and
$A=365$ for the digital-asset settings.

\subsection{Single-Asset Volatility Estimators}

\subsubsection{Naive Volatility Estimator}

The naive estimator uses the uncentered second moment:
\begin{equation}
\widehat\sigma_t
=
\sqrt{
A\cdot\frac{1}{W}
\sum_{i=1}^{W}r_{t-i}^{2}
}.
\label{eq:naive_vol}
\end{equation}
Unlike a sample standard deviation, this quantity includes the drift term.

\subsubsection{Realized Volatility}

For lookback $L=20$,
\begin{equation}
\widehat\sigma_t
=
\sqrt{A}\,
\operatorname{sd}
\left(
 r_{t-L},\ldots,r_{t-1}
\right),
\label{eq:realized_vol}
\end{equation}
where the implementation uses the sample standard deviation ($\mathrm{ddof}=1$).

\subsubsection{EWMA}

With half-life $h=20$,
\begin{equation}
\lambda
=
\exp\!\left(-\frac{\ln 2}{h}\right).
\end{equation}
For observations $s$ inside the trailing estimation window, the implementation
applies the \texttt{adjust=False} EWMA recursion
\begin{equation}
 v_s
 =
 \lambda v_{s-1}
 +(1-\lambda)r_s^2,
\end{equation}
initialized by the first squared return in the supplied window.  Because the
window used at routing time $t$ ends at $t-1$, the decision-time forecast is
\begin{equation}
\widehat\sigma_t
=
\sqrt{A\,v_{t-1}}.
\label{eq:ewma_vol}
\end{equation}
Thus the notation does not imply access to $r_t$ when the action for the next
period is formed.

\subsubsection{Buy-and-Hold Degenerate Estimator}

The implementation defines
\begin{equation}
\widehat\sigma_t\equiv v^\star.
\label{eq:buy_hold_estimator}
\end{equation}
When paired with inverse-volatility scaling, this yields unit exposure.

\subsubsection{AR(1) on Squared Returns}

Within each estimation window, OLS is fit to
\begin{equation}
 h_t
 =
 \alpha+\beta r_{t-1}^{2},
\end{equation}
and the forecast is
\begin{equation}
\widehat\sigma_t
=
\sqrt{
A\cdot\max(h_t,10^{-9})
}.
\label{eq:ar1_vol}
\end{equation}

\subsubsection{AR(2) on Squared Returns}

The two-lag specification is
\begin{equation}
 h_t
 =
 \alpha
 +\beta_1 r_{t-1}^{2}
 +\beta_2 r_{t-2}^{2},
\end{equation}
with
\begin{equation}
\widehat\sigma_t
=
\sqrt{
A\cdot\max(h_t,10^{-9})
}.
\label{eq:ar2_vol}
\end{equation}
Both AR estimators are refit on every window.

\subsubsection{GARCH$(p,q)$}

Returns are internally rescaled as $\widetilde r_t=1000r_t$.  With zero
conditional mean and normal innovations,
\begin{equation}
 h_t
 =
 \omega
 +\sum_{i=1}^{q}\alpha_i\widetilde\varepsilon_{t-i}^{2}
 +\sum_{j=1}^{p}\beta_j h_{t-j}.
\label{eq:garch_state}
\end{equation}
The annualized forecast is
\begin{equation}
\widehat\sigma_t
=
\frac{\sqrt{h_T}}{1000}\sqrt{A}.
\label{eq:garch_forecast}
\end{equation}
Reported defaults use $p=q=1$.  On fit failure or a non-finite forecast,
the implementation falls back to trailing sample volatility.

\subsubsection{GJR-GARCH$(1,1,1)$}

The asymmetric variance recursion is
\begin{equation}
 h_t
 =
 \omega
 +\left(
 \alpha
 +\gamma\mathbf 1[\varepsilon_{t-1}<0]
 \right)\varepsilon_{t-1}^{2}
 +\beta h_{t-1}.
\label{eq:gjr}
\end{equation}
Parameters are recalibrated every 63 steps using a trailing 252-observation
window.  The fallback ladder is
\begin{equation}
\text{GJR forecast}
\rightarrow
\text{EWMA}(h=21)
\rightarrow
\text{realized volatility}.
\end{equation}

\subsubsection{Regime GJR-GARCH}

The local GJR regime is assigned by realized-variance terciles:
\begin{equation}
 \kappa_t^{\mathrm{GJR}}
 =
 \begin{cases}
 \mathrm{low}, & \mathrm{RV}_{t-1}<q_{1/3},\\
 \mathrm{mid}, & q_{1/3}\le\mathrm{RV}_{t-1}<q_{2/3},\\
 \mathrm{high}, & \mathrm{RV}_{t-1}\ge q_{2/3}.
 \end{cases}
\label{eq:regime_gjr_label}
\end{equation}
A separate GJR-GARCH process is calibrated for each regime, and the forecast
uses the parameters indexed by $\kappa_t^{\mathrm{GJR}}$. This symbol is local
to the estimator and is not the Router pair index $k_t$.

\subsubsection{HAR-RV}

Using $\mathrm{RV}_t=r_t^2$, define
\begin{align}
\mathrm{RV}^{(d)}_t
&=\mathrm{RV}_{t-1},\\
\mathrm{RV}^{(w)}_t
&=\frac{1}{5}\sum_{i=1}^{5}\mathrm{RV}_{t-i},\\
\mathrm{RV}^{(m)}_t
&=\frac{1}{22}\sum_{i=1}^{22}\mathrm{RV}_{t-i}.
\end{align}
The HAR model is
\begin{equation}
\mathrm{RV}_t
=
\beta_0
+\beta_d\mathrm{RV}^{(d)}_t
+\beta_w\mathrm{RV}^{(w)}_t
+\beta_m\mathrm{RV}^{(m)}_t
+\epsilon_t.
\label{eq:har_rv}
\end{equation}
Coefficients are estimated by ridge regression with an unpenalized intercept:
\begin{align}
\widehat\beta
&=
\left(X^\top X+\lambda I_0\right)^{-1}X^\top y,\\
I_0&=\operatorname{diag}(0,1,\ldots,1),
\qquad \lambda=10^{-6}.
\label{eq:har_ridge}
\end{align}
The forecast is
\begin{equation}
\widehat\sigma_t
=
\sqrt{A\widehat{\mathrm{RV}}_t}.
\end{equation}

\subsubsection{HAR-RV-Rates}

The implementation augments HAR with a quarterly term and lagged yield-curve
features in log-variance space:
\begin{align}
\ln\mathrm{RV}_t
={}&\beta_0
+\beta_d\mathrm{RV}^{(d)}_t
+\beta_w\mathrm{RV}^{(w)}_t\\
&+\beta_m\mathrm{RV}^{(m)}_t
+\beta_q\mathrm{RV}^{(q)}_t
+\gamma^\top Z_{t-1}
+\epsilon_t.
\label{eq:har_rates}
\end{align}
where
\begin{equation}
Z_t
=
[\mathrm{t10y2y}_t,
  \mathrm{t10y3m}_t,
  \Delta\mathrm{t10y2y}_t,
  \Delta\mathrm{t10y3m}_t]^\top,
\end{equation}
$\mathrm{RV}^{(q)}$ is the 63-day mean, and ridge regularization uses
$\lambda=10^{-4}$.  The forecast is
\begin{equation}
\widehat\sigma_t
=
\sqrt{
A\exp\left(\widehat{\ln\mathrm{RV}_t}\right)
}.
\label{eq:har_rates_forecast}
\end{equation}

\subsubsection{Regime-Specific HAR-Rates}

The regime-specific variant fits independent coefficient vectors
$\beta_{\kappa_t^{\mathrm{HAR}}}$ after constructing the continuous lagged
feature sequence:
\begin{equation}
\ln\mathrm{RV}_t
=
X_t^\top\beta_{\kappa_t^{\mathrm{HAR}}}
+\epsilon_t,
\label{eq:rs_har}
\end{equation}
where the local regime label $\kappa_t^{\mathrm{HAR}}$ is assigned by
realized-variance terciles.

\subsubsection{Lasso Volatility}

Using 22 lags of squared returns, the estimator solves
\begin{equation}
\widehat\beta
=
\arg\min_\beta
\left[
\frac{1}{2n}\|y-X\beta\|_2^2
+\alpha\|\beta\|_1
\right],
\qquad
\alpha=0.1.
\label{eq:lasso_vol}
\end{equation}

\subsubsection{LightGBM Volatility}

The model predicts next-step realized variance from 22 lagged realized-variance
features:
\begin{equation}
\widehat{\mathrm{RV}}_{t+1}
=f_{\theta}^{\mathrm{LGBM}}(x_t).
\label{eq:lgbm_vol}
\end{equation}
The reported implementation uses 50 trees, learning rate 0.05, 31 leaves,
and refits every 21 steps.

\subsubsection{Random-Forest Volatility}

With a five-observation realized-volatility window, define the generic
feature at date $s$ as
\begin{equation}
\mathrm{RV}_s
=
\sqrt{A}\,
\operatorname{sd}
\left(r_{s-4},\ldots,r_s\right).
\end{equation}
At routing time $t$, only lagged RV features are supplied (including the
implemented lags $\{1,5,22\}$), so the feature construction does not imply use
of the yet-unavailable decision-period return.  The prediction is the average
over $B=100$ trees:
\begin{equation}
\widehat{\mathrm{RV}}_{t+1}
=
\frac{1}{B}
\sum_{b=1}^{B}
T_b(x_t),
\label{eq:rf_vol}
\end{equation}
where $x_t$ contains RV lags $\{1,5,22\}$ and optionally a three-state HMM
regime label.

\subsubsection{RNN / MLP Volatility}

The sequence model maps 21 lagged observations into a next-step realized-variance
forecast:
\begin{equation}
\widehat{\mathrm{RV}}_{t+1}
=
f_\theta
\left(
\mathrm{RV}_{t-20:t}
\right).
\label{eq:rnn_vol}
\end{equation}
The default backend is a one-hidden-layer MLP with 32 hidden units; a SimpleRNN
path is available when TensorFlow is installed.

\subsubsection{XGBoost with VIX and Rates}

The XGBoost estimator implements
\begin{equation}
\widehat{\mathrm{RV}}_{t+1}
=
f_{\theta}^{\mathrm{XGB}}(x_t),
\label{eq:xgb_vol}
\end{equation}
where all features are shifted one day and include HAR terms, leverage,
volatility-of-volatility, jumps, VIX variance, changes in VIX variance, yield-curve
features, inversion indicators, and interactions.  In particular,
\begin{equation}
\mathrm{VIXVar}_t
=
\frac{(\mathrm{VIX}_t/100)^2}{A},
\end{equation}
and the variance-risk-premium feature is
\begin{equation}
\mathrm{VRP}_t
=
\frac{
\mathrm{VIXVar}_t
}{
\mathrm{RV}^{(w)}_t+10^{-8}
}.
\label{eq:vrp}
\end{equation}
A jump indicator is
\begin{equation}
J_t
=
\mathbf 1
\left[
|r_t|>3\sigma_{22,t}
\right].
\end{equation}

\subsubsection{Hybrid EWMA Regime}

Two EWMA variance paths are maintained within the supplied trailing
window:
\begin{equation}
 v_s^{(j)}
 =
 \lambda_jv_{s-1}^{(j)}
 +(1-\lambda_j)r_s^2,
\qquad
\lambda_j=(1/2)^{1/h_j},
\end{equation}
with
\begin{equation}
 h_{\mathrm{fast}}=5,
\qquad
 h_{\mathrm{slow}}=40,
\end{equation}
and
\begin{equation}
\sigma_s^{(j)}
=
\sqrt{A v_s^{(j)}}.
\label{eq:hybrid_ewma}
\end{equation}
At routing time $t$, the supplied window ends at $t-1$.  The scalar
\texttt{estimate\_window} path returns $\sigma_{t-1}^{(\mathrm{slow})}$;
the fast path and regime probabilities are exposed through component
diagnostics but do not enter the returned scalar forecast.

\subsubsection{Dynamic Precision Ensemble}

For submodel $m$ over a trailing loss window $W_L=21$,
\begin{equation}
 L_m
 =
 \sum_{i=1}^{W_L}
 \ell
 \left(
 A r_{t-i}^{2},
 \widehat\sigma_{m,t-i}^{2}
 \right).
\end{equation}
The implemented QLIKE loss is
\begin{equation}
\ell_{\mathrm{QLIKE}}(x,f)
=
\ln f+\frac{x}{f},
\label{eq:qlike}
\end{equation}
and the optional MSE loss is
\begin{equation}
\ell_{\mathrm{MSE}}(x,f)
=(x-f)^2.
\end{equation}
Weights are inverse-loss rather than softmax weights:
\begin{equation}
\pi_m
=
\frac{
1/\max(L_m,\varepsilon)
}{
\sum_k1/\max(L_k,\varepsilon)
},
\end{equation}
and
\begin{equation}
\widehat\sigma_t
=
\sum_m\pi_m\widehat\sigma_{m,t}.
\label{eq:dynamic_precision}
\end{equation}

\subsubsection{Intraday Realized Volatility}

With lookback $L=3$ over a precomputed intraday-RV series,
\begin{equation}
\widehat\sigma_t
=
\frac{1}{L}
\sum_{j=1}^{L}
\mathrm{RV}^{\mathrm{intra}}_{t-j}.
\label{eq:intraday_rv}
\end{equation}
The series is shifted one day and falls back to a 20-day return-volatility
estimate when required.

\subsubsection{Range-Based Volatility}

For a completed trading day $s$, the Parkinson variance estimate is
\begin{equation}
\widehat v_s^{\mathrm{Park}}
=
\frac{
\left(\ln(H_s/L_s)\right)^2
}{4\ln2},
\end{equation}
and the default Garman--Klass estimate is
\begin{equation}
\widehat v_s^{\mathrm{GK}}
=
\frac{1}{2}
\left(\ln\frac{H_s}{L_s}\right)^2
-
(2\ln2-1)
\left(\ln\frac{C_s}{O_s}\right)^2.
\label{eq:garman_klass}
\end{equation}
The range series is shifted by one day before routing.  Hence the annualized
20-day forecast formed at $t$ can be written as
\begin{equation}
\widehat\sigma_t
=
\sqrt{
A\cdot\frac{1}{20}
\sum_{j=1}^{20}\widehat v_{t-j}
}.
\label{eq:range_vol}
\end{equation}

\subsubsection{Crypto Composite Volatility}

The composite forecast combines intraday RV, range volatility, and EWMA:
\begin{equation}
\widetilde\sigma_t
=
0.50\widehat\sigma_t^{\mathrm{intra}}
+0.25\widehat\sigma_t^{\mathrm{range}}
+0.25\widehat\sigma_t^{\mathrm{EWMA}},
\end{equation}
and
\begin{equation}
\widehat\sigma_t
=
\operatorname{clip}
\left(
\widetilde\sigma_t,
0.001,
5.0
\right).
\label{eq:crypto_composite}
\end{equation}
Missing component forecasts are replaced by the cross-component median.

\paragraph{Wavelet estimator.}
A wavelet source file exists, but the class is fully commented out and is not an
executable estimator in the effective library.

\subsection{Single-Asset Controllers}

All single-asset controllers map $(v^\star,\widehat\sigma_t,w_{t-1})$ into a
candidate exposure.  The execution engine then clips the resulting exposure to
the task-specific global bounds.  All controllers hold the previous exposure
when the volatility input is missing, non-finite, or non-positive.

\subsubsection{Constant Weight}
\begin{equation}
w_t=c,
\qquad
c=1.
\label{eq:constant_weight}
\end{equation}

\subsubsection{Naive Scaling}
\begin{equation}
w_t^{\mathrm{raw}}
=
\frac{v^\star}
{\max(\widehat\sigma_t,\varepsilon)},
\end{equation}
followed by a hard-coded no-trade band $\delta=0.05$:
\begin{equation}
w_t
=
\begin{cases}
w_{t-1}, & |w_t^{\mathrm{raw}}-w_{t-1}|<\delta,\\
w_t^{\mathrm{raw}}, & \text{otherwise}.
\end{cases}
\label{eq:naive_scaling}
\end{equation}

\subsubsection{Clipped Volatility Targeting}
\begin{equation}
w_t^{\mathrm{raw}}
=
\operatorname{clip}
\left(
\frac{v^\star}{\max(\widehat\sigma_t,\varepsilon)},
 w_{\min},w_{\max}
\right),
\end{equation}
with
\begin{equation}
w_t=w_{t-1}
\quad\text{if}\quad
|w_t^{\mathrm{raw}}-w_{t-1}|<0.05.
\label{eq:vol_target_clip}
\end{equation}
The controller defaults are $w_{\min}=0$ and $w_{\max}=1.5$.

\subsubsection{Hysteresis Controller}

The formula is the same as clipped volatility targeting, but the hold condition is
inclusive:
\begin{equation}
w_t=w_{t-1}
\quad\text{if}\quad
|w_t^{\mathrm{raw}}-w_{t-1}|\le0.05.
\label{eq:hysteresis}
\end{equation}

\subsubsection{Variance Scaling}

Let $\bar\sigma_t$ be the mean of up to the last 252 volatility forecasts:
\begin{equation}
\bar\sigma_t
=
\frac{1}{n}
\sum_{i=1}^{n}
\widehat\sigma_{t-i},
\qquad n\le252.
\end{equation}
The raw exposure uses inverse variance:
\begin{equation}
w_t^{\mathrm{raw}}
=
\frac{v^\star\bar\sigma_t
}{
\max(\widehat\sigma_t^2,\varepsilon)
}.
\label{eq:variance_scaling}
\end{equation}
A 0.05 no-trade band is applied afterward.

\subsubsection{Trend Filter}

Using a 63-day return window, define the controller-local trend score
\begin{equation}
q_t^{\mathrm{trend}}
=
\frac{
\operatorname{mean}(r_{t-63:t-1})
}{
\operatorname{sd}(r_{t-63:t-1})
}.
\end{equation}
For the default smooth gate,
\begin{equation}
m_t^{\mathrm{trend}}
=
\operatorname{clip}
\left(
0.5+0.5\tanh(0.75q_t^{\mathrm{trend}}),
 m_{\mathrm{floor}},
1
\right),
\end{equation}
with $m_{\mathrm{floor}}=0.85$. The exposure is
\begin{equation}
w_t
=
\operatorname{clip}
\left(
 m_t^{\mathrm{trend}}\frac{v^\star}{\widehat\sigma_t},
 w_{\min},w_{\max}
\right).
\label{eq:trend_filter}
\end{equation}
The optional hard gate sets $m_t^{\mathrm{trend}}=1$ for
$q_t^{\mathrm{trend}}>0$ and $m_t^{\mathrm{trend}}=m_{\mathrm{floor}}$
otherwise. These symbols are local to the controller and are distinct from the
Router state $z_t$ and hold/switch decision $g_t$.

\subsubsection{Regime-Switch Controller}

Let
\begin{equation}
\tau_t
=
Q_{0.8}
\left(
\{\widehat\sigma_{t-i}\}_{i=1}^{252}
\right).
\end{equation}
The regime multiplier is
\begin{equation}
m_t
=
\begin{cases}
0.5,&\widehat\sigma_t\ge\tau_t,\\
1.0,&\text{otherwise},
\end{cases}
\end{equation}
and
\begin{equation}
w_t
=
\operatorname{clip}
\left(
 m_t\frac{v^\star}{\widehat\sigma_t},
0,2
\right).
\label{eq:regime_controller}
\end{equation}
Before 20 volatility observations are available, the multiplier defaults to 1.

\subsubsection{Drawdown Brake}

Define drawdown
\begin{equation}
d_t
=
\max\left(
0,
1-\frac{E_t}{E_t^{\mathrm{peak}}}
\right).
\end{equation}
The brake is
\begin{equation}
b_t
=
\begin{cases}
1,&d_t\le d_{\mathrm{start}},\\[2pt]
\max\left(
 b_{\min},
 1-k\frac{d_t-d_{\mathrm{start}}}
 {d_{\mathrm{full}}-d_{\mathrm{start}}}
\right),
&d_t>d_{\mathrm{start}},
\end{cases}
\end{equation}
where $d_{\mathrm{start}}=0.10$, $d_{\mathrm{full}}=0.30$, $k=1$, and
$b_{\min}=0.75$.  Exposure is
\begin{equation}
w_t
=
\operatorname{clip}
\left(
 b_t\frac{v^\star}{\widehat\sigma_t},
 w_{\min},w_{\max}
\right).
\label{eq:drawdown_brake}
\end{equation}

\subsubsection{Drawdown-Modulated Controller}

With
\begin{equation}
d_t
=
\frac{E_t^{\mathrm{peak}}-E_t}{E_t^{\mathrm{peak}}},
\end{equation}
the modulation factor is
\begin{equation}
\phi_t
=
\begin{cases}
1,&d_t\le d_{\mathrm{start}},\\[2pt]
1-\dfrac{d_t-d_{\mathrm{start}}}{d_{\max}-d_{\mathrm{start}}}
\left(1-\phi_{\min}\right),
&d_{\mathrm{start}}<d_t<d_{\max},\\[4pt]
\phi_{\min},&d_t\ge d_{\max},
\end{cases}
\end{equation}
where $d_{\mathrm{start}}=0.10$, $d_{\max}=0.30$, and $\phi_{\min}=0.65$.
Then
\begin{equation}
w_t
=
\operatorname{clip}
\left(
\phi_t\frac{v^\star}{\widehat\sigma_t},
 w_{\min},w_{\max}
\right).
\label{eq:drawdown_modulated}
\end{equation}

\subsubsection{CVaR / Expected-Shortfall Targeting}

Over a 252-day loss buffer $\ell_i=-r_i$ and $\alpha=0.95$,
\begin{equation}
\mathrm{VaR}_\alpha
=
Q_\alpha(\ell),
\end{equation}
and
\begin{equation}
\widehat{\mathrm{ES}}
=
\mathbb E
\left[
\ell\mid\ell\ge\mathrm{VaR}_\alpha
\right].
\end{equation}
The controller targets $\mathrm{ES}^\star=0.02$:
\begin{equation}
w_t
=
\operatorname{clip}
\left(
\frac{\mathrm{ES}^\star}
{\max(\widehat{\mathrm{ES}},\varepsilon)},
 w_{\min},w_{\max}
\right).
\label{eq:es_target}
\end{equation}
This controller does not use $\widehat\sigma_t$ in the exposure rule.

\subsubsection{Priority-Stack Controller}

First compute
\begin{equation}
w_t^{\mathrm{base}}
=
\operatorname{clip}
\left(
\frac{v^\star}{\widehat\sigma_t},
 w_{\min},w_{\max}
\right).
\end{equation}
Three controller-local multipliers are then formed. Here,
$q_t^{\mathrm{trend}}$ is the same 63-day trend score defined in the
Trend-Filter controller above; it is controller-local and is not the Router
state label $z_t$.
\begin{equation}
u_t^{\mathrm{trend}}
=
\begin{cases}
1,&q_t^{\mathrm{trend}}>0,\\
0.80,&\text{otherwise},
\end{cases}
\end{equation}
\begin{equation}
u_t^{\mathrm{dd}}
=
\operatorname{clip}
\left(
1-k\frac{d_t-d_{\mathrm{start}}}{d_{\mathrm{full}}-d_{\mathrm{start}}},
0.80,1
\right),
\end{equation}
and
\begin{equation}
u_t^{\mathrm{tail}}
=
\operatorname{clip}
\left(
\frac{\mathrm{ES}^{\lim}}{\widehat{\mathrm{ES}}},
0.80,1
\right),
\qquad
\mathrm{ES}^{\lim}=0.03.
\end{equation}
The default combination is the mean,
\begin{equation}
G_t
=
\frac{1}{3}
\left(
 u_t^{\mathrm{trend}}
 +u_t^{\mathrm{dd}}
 +u_t^{\mathrm{tail}}
\right),
\label{eq:priority_gate}
\end{equation}
although product and minimum combination modes are also implemented.
The target is
\begin{equation}
\widetilde w_t
=
\operatorname{clip}
\left(
G_t w_t^{\mathrm{base}},
 w_{\min},w_{\max}
\right),
\end{equation}
and the step-limited update is
\begin{equation}
w_t
=
w_{t-1}
+
\operatorname{clip}
\left(
\widetilde w_t-w_{t-1},
-\Delta_{\max},
\Delta_{\max}
\right).
\label{eq:priority_stack}
\end{equation}
Here, $\Delta_{\max}=0.35$.

\subsubsection{Volatility-Shock Throttle}

The base target is modified by a shock indicator:
\begin{equation}
w_t^{\mathrm{base}}
=
\frac{v^\star}{\widehat\sigma_t}
\begin{cases}
\kappa,
&
\widehat\sigma_t
\ge
\mu\operatorname{median}
\left(\{\widehat\sigma\}_{30}\right),\\
1,&\text{otherwise},
\end{cases}
\end{equation}
where $\mu=1.75$ and $\kappa=0.50$.  Exposure changes obey asymmetric limits:
\begin{equation}
\Delta_t
=
\operatorname{clip}
\left(
 w_t^{\mathrm{base}}-w_{t-1},
-0.75,
+0.20
\right),
\end{equation}
and
\begin{equation}
w_t=w_{t-1}+\Delta_t.
\label{eq:shock_throttle}
\end{equation}
A no-trade band of 0.025 is applied.

\subsubsection{Peg-Aware Volatility Controller}

Define the one-period peg deviation
\begin{equation}
p_t
=
|\operatorname{expm1}(r_t)|.
\end{equation}
The peg multiplier is
\begin{equation}
m_t^{\mathrm{peg}}
=
\begin{cases}
1,&p_t\le0.0015,\\
1-\dfrac{p_t-0.0015}{0.0060-0.0015},&0.0015<p_t<0.0060,\\
0,&p_t\ge0.0060,
\end{cases}
\end{equation}
and the drawdown multiplier is
\begin{equation}
m_t^{\mathrm{dd}}
=
\begin{cases}
1,&d_t\le0.02,\\
1-\dfrac{d_t-0.02}{0.08-0.02},&0.02<d_t<0.08,\\
0,&d_t\ge0.08.
\end{cases}
\end{equation}
The exposure update is
\begin{equation}
w_t
=
 w_{t-1}
 +
\operatorname{clip}
\left(
 m_t^{\mathrm{peg}}m_t^{\mathrm{dd}}
 \frac{v^\star}{\widehat\sigma_t}
 -w_{t-1},
-0.25,
0.25
\right).
\label{eq:peg_aware}
\end{equation}

\subsection{Multi-Asset Covariance Estimators}

Let $R\in\mathbb R^{W\times N}$ be the return window.  Every covariance
estimator is annualized and projected onto the positive-semidefinite cone.  For
a symmetric matrix $M$ with eigendecomposition $M=V\Lambda V^\top$,
\begin{equation}
\Pi_{\mathrm{PSD}}(M)
=
V\operatorname{diag}
\left(
\max(\lambda_i,10^{-8})
\right)V^\top,
\label{eq:psd_projection}
\end{equation}
where the matrix is first symmetrized as $(M+M^\top)/2$.

\subsubsection{Sample Covariance}
\begin{equation}
\widehat\Sigma_t
=
A\operatorname{cov}(R).
\label{eq:sample_cov}
\end{equation}

\subsubsection{Expanding Covariance}
\begin{equation}
\widehat\Sigma_t
=
A\operatorname{cov}
\left(
R_{1:t-1}
\right),
\label{eq:expanding_cov}
\end{equation}
with a trailing-window fallback when fewer than 63 historical rows are available.

\subsubsection{EWMA Covariance}

With half-life $h=21$,
\begin{equation}
\delta=(1/2)^{1/h},
\qquad
\omega_i=\delta^{n-i},
\qquad
\widetilde\omega_i
=
\frac{\omega_i}{\sum_j\omega_j},
\end{equation}
and
\begin{equation}
\widehat\Sigma_t
=
A
\sum_i
\widetilde\omega_i
(r_i-\bar r_\omega)
(r_i-\bar r_\omega)^\top.
\label{eq:ewma_cov}
\end{equation}

\subsubsection{Diagonal EWMA Covariance}
\begin{equation}
\widehat\Sigma_t
=
\operatorname{diag}
\left(
\operatorname{diag}
\left(
\widehat\Sigma_t^{\mathrm{EWMA}}
\right)
\right).
\label{eq:diag_ewma_cov}
\end{equation}

\subsubsection{Rolling-Correlation / EWMA-Volatility Covariance}

Let $C_t$ be the rolling correlation matrix and
$\boldsymbol\sigma_t$ the vector of EWMA marginal volatilities.  Then
\begin{equation}
\widehat\Sigma_t
=
C_t\odot
\boldsymbol\sigma_t\boldsymbol\sigma_t^\top.
\label{eq:rolling_corr_cov}
\end{equation}
The correlation window is at most 126 observations.

\subsubsection{Shrunk Sample Covariance}
\begin{equation}
\widehat\Sigma_t
=
(1-\delta)S_t
+\delta\operatorname{diag}(S_t),
\qquad
\delta=0.25.
\label{eq:shrunk_cov}
\end{equation}

\subsubsection{Ledoit--Wolf Covariance}

The implementation uses Ledoit--Wolf shrinkage. On failure, it falls back
to diagonal shrinkage with intensity 0.35:
\begin{equation}
\widehat\Sigma_t
=
(1-0.35)S_t
+0.35\operatorname{diag}(S_t).
\label{eq:lw_fallback}
\end{equation}

\subsubsection{Downside Covariance}

Centered downside returns are
\begin{equation}
\widetilde R
=
\min(R-\bar R,0),
\end{equation}
and
\begin{equation}
\widehat\Sigma_t
=
(1-b)
\operatorname{cov}(\widetilde R)
+b\operatorname{cov}(R),
\qquad
b=0.35.
\label{eq:downside_cov}
\end{equation}

\subsubsection{Robust Median Covariance}

For each series,
\begin{equation}
z
=
\operatorname{clip}
\left(
\frac{R-\operatorname{med}(R)}
{1.4826\operatorname{MAD}(R)},
-4,
4
\right).
\label{eq:robust_z}
\end{equation}
The clipped standardized series is rescaled and its covariance is computed.

\subsubsection{Regime-Switching Covariance}

Let $\Sigma_t^{f}$ and $\Sigma_t^{s}$ denote EWMA covariance matrices with
half-lives 10 and 63.  Then
\begin{equation}
\widehat\Sigma_t
=
\alpha_t\Sigma_t^{f}
+(1-\alpha_t)\Sigma_t^{s},
\end{equation}
where
\begin{equation}
\alpha_t
=
\begin{cases}
0.75,&\mathrm{RV}_t\ge0.18,\\
0.35,&\mathrm{RV}_t<0.18.
\end{cases}
\label{eq:regime_cov}
\end{equation}

\subsubsection{VIX-Scaled Covariance}

With a 252-day VIX median, define the covariance scale
\begin{equation}
\eta_t^{\mathrm{VIX}}
=
\operatorname{clip}
\left[
\left(
\frac{\mathrm{VIX}_t}
{\operatorname{med}_{252}(\mathrm{VIX})}
\right)^2,
0.35,
4
\right],
\end{equation}
and
\begin{equation}
\widehat\Sigma_t
=
\eta_t^{\mathrm{VIX}}\widehat\Sigma_t^{\mathrm{EWMA}}.
\label{eq:vix_cov}
\end{equation}

\subsubsection{PCA Covariance}

Let $S=V\Lambda V^\top$ and retain the top $k=3$ eigenpairs:
\begin{equation}
\widehat\Sigma_t
=
V_k\Lambda_kV_k^\top
+0.15\operatorname{diag}(S).
\label{eq:pca_cov}
\end{equation}

\subsubsection{Dynamic Blend Covariance}

Define the local covariance-stress score
\begin{equation}
\chi_t
=
\operatorname{clip}
\left(
\frac{\mathrm{vol}_{21,t}-0.08}{0.20},
0,
1
\right).
\end{equation}
The covariance blend is
\begin{align}
\widehat\Sigma_t
={}&(0.45-0.20\chi_t)\Sigma_t^{\mathrm{EWMA}}
+0.35\Sigma_t^{\mathrm{LW}}\\
&+(0.20+0.20\chi_t)\Sigma_t^{\mathrm{down}}.
\label{eq:dynamic_blend_cov}
\end{align}

\subsection{Multi-Asset Portfolio Controllers}

Every portfolio controller, except buy-and-hold, passes raw weights through a
shared normalization, cap, and volatility-scaling pipeline.

\paragraph{Long-only normalization.}
\begin{equation}
w_i
\leftarrow
\frac{\max(w_i,0)}
{\sum_j\max(w_j,0)}.
\label{eq:ma_normalize}
\end{equation}
If the denominator is non-positive, the implementation falls back to equal
weights.

\paragraph{Single-name cap.}
With $\bar w=0.45$,
\begin{equation}
w_i\leftarrow\min(w_i,\bar w),
\end{equation}
and the remaining mass is redistributed iteratively over assets with available
capacity.

\paragraph{Portfolio volatility scaling.}
\begin{equation}
\sigma_{p,t}
=
\sqrt{
\mathbf w^\top\widehat\Sigma_t\mathbf w
},
\end{equation}
and
\begin{equation}
\mathbf w_t
=
\mathbf w
\cdot
\operatorname{clip}
\left(
\frac{v^\star}{\max(\sigma_{p,t},10^{-8})},
0,
\Gamma
\right),
\qquad
\Gamma=1.5.
\label{eq:ma_vol_scale}
\end{equation}

\subsubsection{Equal Weight}
\begin{equation}
w_{i,t}=\frac{1}{N}.
\label{eq:eq_weight}
\end{equation}

\subsubsection{Buy-and-Hold}

Once initialized, the controller returns
\begin{equation}
\mathbf w_t=\mathbf w_{t-1},
\label{eq:ma_buy_hold}
\end{equation}
and bypasses the shared cap/volatility-scaling pipeline.

\subsubsection{Inverse Volatility}
\begin{equation}
\widetilde w_i
\propto
\frac{1}{\sigma_i},
\qquad
\sigma_i=\sqrt{\widehat\Sigma_{ii}}.
\label{eq:inverse_vol}
\end{equation}

\subsubsection{Minimum Variance}
\begin{equation}
\widetilde{\mathbf w}
\propto
\widehat\Sigma^{+}\mathbf 1,
\label{eq:min_var}
\end{equation}
where $+$ denotes the Moore--Penrose pseudo-inverse.

\subsubsection{Vol-Capped Minimum Variance}

Starting from minimum-variance weights, assets with marginal volatility
above $c=0.30$ are rescaled as
\begin{equation}
\widetilde w_i
\leftarrow
\widetilde w_i
\frac{c}{\sigma_i},
\qquad
\sigma_i>c.
\label{eq:vol_capped_mv}
\end{equation}

\subsubsection{Equal Risk Contribution}

Initialize $w_i\propto1/\sqrt{\widehat\Sigma_{ii}}$.  At each fixed-point iteration,
\begin{equation}
\mathrm{RC}_i
=
 w_i(\widehat\Sigma\mathbf w)_i,
\end{equation}
and
\begin{equation}
w_i
\leftarrow
\operatorname{normalize}
\left[
 w_i
 \sqrt{
 \frac{\overline{\mathrm{RC}}}
 {\max(\mathrm{RC}_i,10^{-10})}
 }
\right].
\label{eq:erc_iteration}
\end{equation}
The implementation uses 80 iterations.  At convergence,
$\mathrm{RC}_i\approx\overline{\mathrm{RC}}$.

\subsubsection{Diversified Risk Parity}

Let
\begin{equation}
\bar\rho_i
=
\frac{1}{N}
\sum_j
|\rho_{ij}|.
\end{equation}
Starting from ERC weights,
\begin{equation}
\widetilde w_i
=
\frac{w_i^{\mathrm{ERC}}}
{\max(\bar\rho_i,0.25)}.
\label{eq:div_risk_parity}
\end{equation}

\subsubsection{Momentum Tilt}

Define 63-day momentum
\begin{equation}
m_i
=
\sum_{j=1}^{63}r_{i,t-j}.
\end{equation}
Let $q_i$ be its cross-sectional percentile rank.  Starting from inverse-volatility
weights,
\begin{equation}
\widetilde w_i
=
w_i^{\mathrm{IV}}
\max
\left[
0.25,
(q_i-0.5)\theta+1
\right],
\qquad
\theta=0.8.
\label{eq:momentum_tilt}
\end{equation}

\subsubsection{Mean--Variance}

With a 126-day mean-return estimate,
\begin{equation}
\widehat\mu_t
=
252\cdot
\operatorname{mean}_{126}(r),
\end{equation}
and risk aversion $\gamma=6$,
\begin{equation}
\widetilde{\mathbf w}
\propto
\frac{1}{\gamma}
\widehat\Sigma^{+}\widehat\mu_t.
\label{eq:mean_variance}
\end{equation}

\subsubsection{Regime-Aware Risk Budget}

Start from ERC weights.  If 21-day portfolio volatility exceeds 0.18, multiply the
hard-coded defensive assets $\{\mathrm{IEF},\mathrm{TLT},\mathrm{GLD},\mathrm{UUP}\}$ by
1.6 before normalization:
\begin{equation}
\widetilde w_i
=
\begin{cases}
1.6w_i^{\mathrm{ERC}},
&i\in\mathcal D\ \text{and}\ \mathrm{vol}_{21}>0.18,\\
w_i^{\mathrm{ERC}},&\text{otherwise},
\end{cases}
\label{eq:regime_risk_budget}
\end{equation}
where $\mathcal D=\{\mathrm{IEF},\mathrm{TLT},\mathrm{GLD},\mathrm{UUP}\}$.

\subsubsection{Portfolio Drawdown Brake}

Starting from inverse-volatility weights, the controller applies
\begin{equation}
\widetilde{\mathbf w}_t
=
\begin{cases}
0.55\mathbf w_t^{\mathrm{IV}},
&\text{proxy drawdown}<-0.08,\\
\mathbf w_t^{\mathrm{IV}},&\text{otherwise}.
\end{cases}
\label{eq:portfolio_dd_brake}
\end{equation}

\subsubsection{Portfolio Hysteresis}

Let $\mathbf w_t^{\mathrm{cand}}$ be the post-processed candidate weight vector.  Then
\begin{equation}
\mathbf w_t
=
\begin{cases}
\mathbf w_{t-1},
&
\|\mathbf w_t^{\mathrm{cand}}-\mathbf w_{t-1}\|_1<0.05,\\
\mathbf w_t^{\mathrm{cand}},&\text{otherwise}.
\end{cases}
\label{eq:portfolio_hysteresis}
\end{equation}

\subsection{Execution Equations}

\subsubsection{Single-Asset Execution}

The selected estimator/controller pair obeys
\begin{equation}
\widehat\sigma_t
=
\mathcal E
\left(
 r^{\mathrm{clean}}_{t-W:t-1}
\right),
\end{equation}
\begin{equation}
w_t
=
\begin{cases}
\operatorname{clip}
\left(
\mathcal C(v^\star,\widehat\sigma_t,w_{t-1}),
 w_{\min},w_{\max}
\right),
&t\in\mathcal R,\\
w_{t-1},&\text{otherwise},
\end{cases}
\end{equation}
where $\mathcal R$ denotes rebalance dates.  Turnover and cost are
\begin{equation}
\tau_t
=
|w_t-w_{t-1}|,
\qquad
c_t
=
\tau_t\frac{\mathrm{bps}}{10^4},
\end{equation}
and next-period strategy return and equity are
\begin{equation}
r_{t+1}^{\mathrm{strat}}
=
 w_t r_{t+1}-c_t,
\end{equation}
\begin{equation}
E_{t+1}
=
E_t\exp
\left(
 r_{t+1}^{\mathrm{strat}}
\right).
\label{eq:single_execution}
\end{equation}

\subsubsection{Multi-Asset Execution}

The selected covariance estimator and portfolio controller generate
\begin{equation}
\widehat\Sigma_t
=
\mathcal E
\left(
R_{t-W:t-1}
\right),
\end{equation}
\begin{equation}
\mathbf w_t
=
\mathcal C
\left(
 v^\star,
 \widehat\Sigma_t,
 R_{t-W:t-1},
 \mathbf w_{t-1}
\right).
\end{equation}
Gross exposure, cash, and turnover are
\begin{equation}
G_t^{\mathrm{gross}}
=
\sum_i|w_{t,i}|,
\qquad
w_t^{\mathrm{cash}}=1-G_t^{\mathrm{gross}},
\end{equation}
\begin{equation}
\tau_t
=
\sum_i|w_{t,i}-w_{t-1,i}|.
\end{equation}
The next-period strategy return is
\begin{equation}
r_{t+1}^{\mathrm{strat}}
=
\mathbf w_t^\top\mathbf r_{t+1}
+w_t^{\mathrm{cash}}r_{t+1}^{f}
-
\tau_t\frac{\mathrm{bps}}{10^4}.
\label{eq:multi_execution}
\end{equation}
The risk-free leg is
\begin{equation}
r_t^f
=
\ln
\left(
1+
\frac{\mathrm{DGS3MO}_t/100}{252}
\right).
\label{eq:risk_free_leg}
\end{equation}

\subsection{Router Scoring Function}

For candidate pair $k$, the deterministic router score is
\begin{align}
\mathrm{Score}_t(k)
={}&
\underbrace{
\pi
\left(
\mathrm{SR}_k
-\lambda_{dd}d_k^+
\right)
}_{\mathrm{performance}}
+
\underbrace{
\beta_{\mathrm{reg}}B_t(k)
}_{\mathrm{regime\ bias}}
\nonumber\\
&-
\underbrace{
\left(
\lambda_{\mathrm{inv}}\iota_k
+
\lambda_{\mathrm{exc}}\chi_k
\right)
}_{\mathrm{diagnostics}}
-
\underbrace{
\lambda_{\mathrm{sw}}
\mathbf 1[k\neq k_{t-1}]
}_{\mathrm{switch\ penalty}}.
\label{eq:app_router_score}
\end{align}
The regime-bias term is
\begin{align}
B_t(k)
={}&b_t^{\mathrm{pair}}(k)
+b_t^{\mathrm{est}}(k)
+b_t^{\mathrm{ctrl}}(k)\\
&+
\mathbf 1\!\left[
 b_t^{\mathrm{pair}}
+b_t^{\mathrm{est}}
+b_t^{\mathrm{ctrl}}
=0
\right]
 b_t^{\mathrm{heur}}(k).
\label{eq:router_bias}
\end{align}
Reported base defaults are
\begin{align}
\pi&=1,
&\beta_{\mathrm{reg}}&=1,
&\lambda_{dd}&=0.5,\\
\lambda_{\mathrm{inv}}&=2,
&\lambda_{\mathrm{exc}}&=1,
&\lambda_{\mathrm{sw}}&=0.
\label{eq:router_defaults}
\end{align}
The AI regime router overrides $\beta_{\mathrm{reg}}$ to $2.5$.
The performance term is set to zero when the candidate has fewer than the
minimum required observations.  In the audited evaluation path, the
invalid-rate and exception-rate diagnostics are not populated, so the
corresponding penalty is zero in the reported runs.

\subsection{Setting-Specific Constants}

\begin{table*}[t]
\centering
\caption{Core execution constants by evaluation setting.}
\label{tab:math_constants_by_setting}
\small
\begin{tabular}{lcccc}
\toprule
& \textbf{S\&P 500}
& \textbf{Digital Market (Bitcoin)}
& \textbf{USDT}
& \textbf{Multi-Asset} \\
\midrule
Target volatility $v^\star$ & 0.10 & 0.35 & 0.02 & 0.10 \\
Estimation window $W$ & 252 & 90 & 90 & 126 \\
Transaction cost (bps) & 5.0 & 8.0 & 2.0 & 5.0 \\
Exposure bound & $[0,1.5]$ & $[0,1.25]$ & $[0,1.25]$ & cap 0.45 / gross 1.5 \\
Annualization $A$ & 252 & 365 & 365 & 252 \\
\bottomrule
\end{tabular}
\end{table*}

\subsection{Implementation-Specific Deviations}

The following implementation details are important when interpreting the
mathematical definitions above:
\begin{enumerate}
    \item \texttt{NaiveScaling} ignores its constructor parameter object;
    the 0.05 no-trade band is hard-coded.
    \item \texttt{NaiveVolEstimator} uses the uncentered second moment rather
    than the sample variance.
    \item the wavelet estimator is non-functional because its class is fully
    commented out.
    \item \texttt{HybridEWMARegime} returns the slow EWMA scalar on the main
    estimation path; the fast path and regime probabilities are diagnostic
    components only.
    \item \texttt{CVaRESTargeting} ignores the volatility forecast and targets
    expected shortfall directly.
    \item \texttt{DynamicPrecisionEnsemble} uses inverse-loss weighting rather
    than exponential or softmax weighting.
    \item the regime-GJR and regime-HAR estimators use realized-volatility
    quantiles rather than a latent HMM to define their regimes.
    \item GJR-GARCH parameters are held fixed between 63-step recalibrations.
    \item the multi-asset buy-and-hold controller bypasses the common cap and
    volatility-scaling pipeline.
    \item the multi-asset regime-aware risk-budget controller hard-codes
    $\{\mathrm{IEF},\mathrm{TLT},\mathrm{GLD},\mathrm{UUP}\}$ as defensive assets.
    \item missing returns in the multi-asset engine are filled with zero before
    covariance estimation.
    \item the deterministic router's invalid-rate and exception-rate penalties
    are zero in the audited reported evaluation path because those diagnostics
    are not populated by that driver.
\end{enumerate}

\section{Prompt and Routing Decision Format}
\label{app:prompt_routing_format}

This section provides a normalized, implementation-faithful specification of
the prompt interfaces used by
\textsc{VolRouter}.  The boxes below preserve the audited decision roles,
constraints, admissible labels, and output schemas; they are \emph{not claimed
to be byte-for-byte transcriptions of the source constants}. The purpose is
to specify the decision interface and admissible outputs rather than expose
provider-specific prompt boilerplate.
Rather than asking a language model to directly generate portfolio weights,
\textsc{VolRouter} decomposes routing into structured decisions over a fixed
library of estimator--controller pairs.
The language model is used only to infer the current market state,
determine whether the active pair should be retained, and, when necessary,
select a replacement from an explicitly supplied candidate set.

The prompted decision process follows three stages:
\[
\text{State Inference}
\rightarrow
\text{Switch Review}
\rightarrow
\text{Selection}.
\]

The single-asset and multi-asset implementations use the same general
structure, but differ in the state representation and regime labels.
All reported routing runs use structured JSON outputs so that the language
model cannot directly modify estimator parameters, controller parameters,
portfolio constraints, or the policy library.

\subsection{Single-Asset Market-State Prompt}

For single-asset volatility targeting, the first prompt summarizes the
current market into one of three volatility regimes:
\[
z_t \in
\{\texttt{low},\texttt{middle},\texttt{high}\}.
\]

The input contains only market information available at the decision time.
It does not provide future returns or candidate-pair rankings.

\begin{routerpromptbox}{Single-Asset Volatility-Regime Prompt}
You are the market-state inference component of a volatility-targeting\par
router.\par
\promptgap
Your task is to classify the current volatility environment using only\par
the market information supplied below.\par
\promptgap
Choose exactly one regime:\par
\promptgap
- low\par
- middle\par
- high\par
\promptgap
Use only information available at the current decision time.\par
Do not infer or use future returns.\par
Do not use future strategy performance.\par
Do not rank estimator--controller pairs in this step.\par
\promptgap
Return a JSON object with the following format:\par
\promptgap
\{\par
  "vol\_regime": "<low | middle | high>",\par
  "confidence": "<confidence score>",\par
  "reason": "<brief explanation>"\par
\}\par
\promptgap
Keep the explanation shorter than 24 words and base the classification only on the\par
provided market evidence.\par
\end{routerpromptbox}

The resulting regime is represented as
\[
z_t =
\mathcal{R}_{\theta}
\left(
\mathcal I_t^{\mathrm{market}}
\right).
\]

The regime label is subsequently passed to the switch-review and
pair-selection layers as contextual information.

\subsection{Single-Asset Switch-Review Prompt}

The second prompt decides whether to retain the active policy pair.
It does not select a replacement; it only returns a hold/switch decision.

\begin{routerpromptbox}{Single-Asset Switch-Review Prompt}
You are the switch-review component of a volatility-targeting router.\par
\promptgap
A single estimator--controller pair is currently active.\par
Using the supplied market state and recent routing evidence, decide whether\par
the current pair should be retained or whether the router should consider a\par
replacement.\par
\promptgap
Available evidence may include:\par
\promptgap
- current volatility regime,\par
- recent performance of the active pair,\par
- recent relative rankings of candidate pairs,\par
- recent risk-control behavior,\par
- current routing state.\par
\promptgap
Choose exactly one action:\par
\promptgap
- hold\par
- switch\par
\promptgap
Return only a JSON object in the following format:\par
\promptgap
\{\par
  "action": "<hold | switch>"\par
\}\par
\promptgap
Choose "hold" when the current pair remains appropriate for the observed\par
state and there is insufficient evidence for replacement.\par
\promptgap
Choose "switch" only when the supplied evidence indicates that another\par
candidate is more appropriate for the current market state.\par
\end{routerpromptbox}

Formally,
\[
g_t =
\mathcal{G}_{\theta}
\left(
z_t,
\mathcal I_t^{\mathrm{routing}},
k_{t-1}
\right)
\in
\{\mathrm{hold},\mathrm{switch}\}.
\]

If
\[
g_t=\mathrm{hold},
\]
then
\[
k_t=k_{t-1},
\]
and no pair-selection prompt is called.

\subsection{Single-Asset Pair-Selection Prompt}

Pair selection is invoked only after the switch-review layer has decided
that the active pair should not be retained.
The active pair is excluded from the candidate set supplied to this prompt.

\begin{routerpromptbox}{Single-Asset Pair-Selection Prompt}
You are the pair-selection component of a volatility-targeting router.\par
\promptgap
The switch-review layer has already decided that the current pair should\par
not be retained.\par
\promptgap
Select exactly one estimator--controller pair from the candidate list\par
provided below.\par
\promptgap
When selecting a pair, consider:\par
\promptgap
- the current volatility regime,\par
- recent candidate performance,\par
- recent relative rankings,\par
- volatility-control quality,\par
- drawdown behavior,\par
- turnover behavior,\par
- whether the candidate is suitable for the current market state.\par
\promptgap
You may select only a pair explicitly included in the candidate list.\par
Do not invent a new estimator, controller, or pair.\par
Do not modify any candidate parameters.\par
\promptgap
Return only:\par
\promptgap
\{\par
  "pair": "<candidate pair name>"\par
\}\par
\end{routerpromptbox}

The routing decision is therefore
\[
k_t
=
\mathcal{S}_{\theta}
\left(
z_t,
\mathcal I_t^{\mathrm{routing}},
\mathcal C_t
\setminus
\{k_{t-1}\}
\right),
\]
where $\mathcal C_t$ is the candidate set supplied at time $t$.

\subsection{Multi-Asset Portfolio-Regime Prompt}

The multi-asset router uses a portfolio-level state representation.
Its portfolio regime belongs to
\[
z_t^{p}
\in
\{
\texttt{risk\_on},
\texttt{balanced},
\texttt{defensive}
\}.
\]

\begin{routerpromptbox}{Multi-Asset Portfolio-Regime Prompt}
You are the portfolio-state inference component of a multi-asset\par
volatility-targeting router.\par
\promptgap
Using the supplied portfolio and market information, classify the current\par
environment into exactly one of the following states:\par
\promptgap
- risk\_on\par
- balanced\par
- defensive\par
\promptgap
The classification should reflect the current risk environment rather\par
than short-term speculation about future returns.\par
\promptgap
Consider only information supplied in the prompt and available at the\par
current decision time.\par
\promptgap
Return:\par
\promptgap
\{\par
  "portfolio\_regime":\par
      "<risk\_on | balanced | defensive>",\par
  "confidence": "<confidence score>",\par
  "reason": "<brief explanation>"\par
\}\par
\end{routerpromptbox}

This state provides portfolio-level context for subsequent routing
decisions.

\subsection{Multi-Asset Market-Only Volatility Prompt}

The implementation also contains a market-only version of the same
multi-asset state-inference interface. It should not be read as an additional
fourth routing stage: both templates return the same
\texttt{risk\_on}/\texttt{balanced}/\texttt{defensive} state space, but differ
in which decision-time state fields are exposed. The market-only template is
separated from candidate performance: it receives cross-asset market
information but no pair rankings or pair-level metrics.

\begin{routerpromptbox}{Multi-Asset Market-Only Regime Prompt}
You are assessing the current cross-asset risk environment.\par
\promptgap
Use only the supplied market variables, which may include:\par
\promptgap
- recent cross-asset returns,\par
- equal-weight portfolio realized volatility,\par
- VIX information,\par
- term-spread information,\par
- credit-spread information.\par
\promptgap
Do not use:\par
\promptgap
- estimator--controller pair rankings,\par
- pair-level returns,\par
- pair-level Sharpe ratios,\par
- future returns,\par
- future portfolio outcomes.\par
\promptgap
Classify the current environment as exactly one of:\par
\promptgap
- risk\_on\par
- balanced\par
- defensive\par
\promptgap
Return:\par
\promptgap
\{\par
  "portfolio\_regime":\par
      "<risk\_on | balanced | defensive>",\par
  "confidence": "<confidence score>",\par
  "reason": "<brief explanation>"\par
\}\par
\end{routerpromptbox}

Separating market-state inference from candidate evaluation prevents the
regime label from being defined retrospectively by whichever policy
performed best.

\subsection{Multi-Asset Switch-Review Prompt}

The multi-asset switch layer decides whether accumulated evidence is
sufficient to replace the active portfolio policy.
The implementation encourages persistent policies rather than
high-frequency strategy switching.

\begin{routerpromptbox}{Multi-Asset Switch-Review Prompt}
You are the switch-review component of a multi-asset portfolio router.\par
\promptgap
Determine whether the currently active estimator--controller pair should\par
be held or replaced.\par
\promptgap
Base the decision on the supplied evidence, including:\par
\promptgap
- current portfolio regime,\par
- recent performance of the active pair,\par
- recent relative candidate rankings,\par
- drawdown behavior,\par
- volatility-control quality,\par
- recent holding duration,\par
- evidence that alternative pairs consistently outperform the active pair.\par
\promptgap
Avoid unnecessary switching.\par
The intended behavior is persistent routing rather than frequent reaction\par
to isolated short-term fluctuations.\par
\promptgap
Use a forward decision horizon of approximately ten trading days.\par
The risk-control objective includes avoiding persistent absolute drawdowns\par
around the approximately 6\% drawdown guardrail used by this switch prompt.\par
This guardrail is distinct from the 10\% portfolio volatility target.\par
\promptgap
Return exactly:\par
\promptgap
\{\par
  "action": "<hold | switch>"\par
\}\par
\end{routerpromptbox}

In the reported implementation, the switch decision is reviewed only at
configured checkpoints rather than every trading day.

\subsection{Multi-Asset Pair-Selection Prompt}

If the switch layer chooses \texttt{switch}, the router receives a
restricted candidate list.
The active pair is excluded, so selection necessarily corresponds to a
change in policy.

\begin{routerpromptbox}{Multi-Asset Pair-Selection Prompt}
You are selecting a replacement estimator--controller pair for a\par
multi-asset volatility-targeting portfolio.\par
\promptgap
The switch-review layer has already determined that the current pair\par
should not be held.\par
\promptgap
Choose exactly one pair from the supplied candidate list.\par
\promptgap
Use the provided evidence to identify the candidate with the strongest\par
expected portfolio behavior over the next decision horizon, while\par
respecting the volatility-control objective.\par
\promptgap
Consider:\par
\promptgap
- current portfolio regime,\par
- recent candidate performance,\par
- short- and medium-horizon candidate rankings,\par
- volatility-control quality,\par
- drawdown behavior,\par
- turnover behavior,\par
- consistency across recent evaluation windows.\par
\promptgap
You may choose only from the supplied candidates.\par
Do not invent a new pair.\par
Do not modify estimator or controller parameters.\par
Do not return the currently active pair.\par
\promptgap
Return only:\par
\promptgap
\{\par
  "pair": "<candidate pair name>"\par
\}\par
\end{routerpromptbox}

The phrase ``expected portfolio behavior'' in this prompt denotes a comparative
choice from supplied historical diagnostics; the model is not asked to produce a
numerical return forecast.

Thus, selection remains a discrete decision over the predefined policy
library:
\[
k_t
\in
\mathcal C_t.
\]

The language model never outputs portfolio weights directly.

\subsection{Switch-Sensitivity Guidance}

The router supports a sensitivity parameter controlling the amount of
evidence required before switching.
For the single-asset implementation this is primarily expressed through
prompt guidance, whereas the multi-asset implementation additionally
enforces minimum-hold and evidence-count requirements in code.

A representative guidance block is:

\begin{routerpromptbox}{Switch-Sensitivity Guidance}
Routing sensitivity determines how readily the active pair should be\par
replaced.\par
\promptgap
Very Low:\par
Prefer strong persistence. Hold the active pair for long periods unless\par
multiple recent evaluation windows consistently indicate that a group of\par
alternatives is superior.\par
\promptgap
Low:\par
Prefer persistence and require sustained evidence before switching.\par
\promptgap
Medium:\par
Balance persistence with responsiveness. Switch when several independent\par
signals consistently favor alternative candidates.\par
\promptgap
High:\par
Respond more quickly when a non-active candidate demonstrates superior\par
recent behavior.\par
\promptgap
Very High:\par
Allow aggressive switching when short-horizon evidence indicates that\par
another candidate is currently better suited to the observed market state.\par
\end{routerpromptbox}

In the multi-asset implementation, these qualitative levels correspond to
explicit gates:

\[
\begin{array}{lccc}
\toprule
\text{Level} & \text{Min. hold} & \text{Better} & \text{Windows}\\
\midrule
\text{Very High} & 0  & 0 & 1\\
\text{High}      & 0  & 0 & 1\\
\text{Medium}    & 30 & 2 & 2\\
\text{Low}       & 60 & 2 & 2\\
\text{Very Low}  & 90 & 3 & 2\\
\bottomrule
\end{array}
\]

This distinction is important because single-asset sensitivity changes
the language-model instruction, while multi-asset sensitivity also
modifies deterministic routing constraints.

\subsection{API and Inference Configuration}

Reported LLM routing uses hosted inference with temperature $0.0$ and
structured JSON outputs. Figure~\ref{fig:router_models} compares the four
backbones labeled in that figure under the corresponding experiment-specific
routing configurations. Review intervals and candidate-top-$N$ values vary by
setting; the S\&P backbone sweep uses a precomputed regime series held fixed
across models, so that comparison isolates switch review and pair selection
rather than end-to-end state inference.

\subsection{Structured Router Output}

The language-model interface is intentionally narrow.
The router expects one of three structured output schemas.

\begin{routerpromptbox}{Single-Asset Regime Output Schema}
\{\par
  "vol\_regime": "<low | middle | high>",\par
  "confidence": "<confidence score>",\par
  "reason": "<brief explanation>"\par
\}\par
\end{routerpromptbox}

\begin{routerpromptbox}{Multi-Asset Regime Output Schema}
\{\par
  "portfolio\_regime": "<risk\_on | balanced | defensive>",\par
  "confidence": "<confidence score>",\par
  "reason": "<brief explanation>"\par
\}\par
\end{routerpromptbox}

\begin{routerpromptbox}{Switch Output Schema}
\{\par
  "action": "<hold | switch>"\par
\}\par
\end{routerpromptbox}

\begin{routerpromptbox}{Pair-Selection Output Schema}
\{\par
  "pair": "<candidate pair identifier>"\par
\}\par
\end{routerpromptbox}

This structured interface prevents unconstrained language generation from
directly changing the portfolio policy.
The LLM can select among available policies but cannot modify the policy
library itself.

\subsection{Parsing, Failure Handling, and Deterministic Fallback}

Router outputs are parsed as structured JSON.
For the single-asset router, parsing proceeds through several recovery
stages:

\begin{enumerate}
    \item direct JSON parsing;
    \item extraction and parsing of a JSON-like object;
    \item recovery of a valid hold/switch action from a structured text
    pattern;
    \item recovery from recognizable natural-language action text; and
    \item deterministic fallback if parsing or API execution fails.
\end{enumerate}

The multi-asset router uses a stricter parser and does not implement all
of the natural-language recovery stages available in the single-asset
router.

Let
$N_{\mathrm{fallback}}$ denote the number of routing decisions ultimately
resolved through a deterministic fallback and
$N_{\mathrm{decision}}$ the total number of prompted routing decisions.
For audit purposes, define

\[
\mathrm{FallbackRate}
=
\frac{
N_{\mathrm{fallback}}
}{
N_{\mathrm{decision}}
}.
\]

Raw model responses are stored during evaluation so that prompt failures
and fallback decisions can be audited independently of portfolio
performance.

\subsection{Prompt-Level Constraints}

Across all prompted layers, the following constraints are enforced by the
routing interface:

\begin{enumerate}
    \item \textbf{No direct portfolio generation.}
    The language model never outputs a portfolio weight or leverage level.

    \item \textbf{Closed candidate set.}
    The pair-selection layer can choose only from the candidates supplied
    by the routing system.

    \item \textbf{No parameter modification.}
    Estimator and controller parameters are fixed before the prompted
    decision.

    \item \textbf{Separated switch and selection decisions.}
    Candidate selection is invoked only after an explicit switch decision.

    \item \textbf{No future information.}
    Market-state prompts use only information available at the decision
    date.

    \item \textbf{Market-state / pair-performance separation.}
    The market-only multi-asset regime prompt explicitly excludes
    candidate-pair rankings and performance metrics.

    \item \textbf{Machine-readable outputs.}
    The prompted decisions are represented as compact JSON records and are
    validated before execution.
\end{enumerate}

These constraints make the language model a policy-selection component
rather than an unconstrained trading agent.
The portfolio action remains generated by the selected, predefined
estimator--controller pair.

\section{Evaluation Protocol}
\label{app:protocol}

\subsection{Generic Walk-Forward Driver}

The generic protocol supports tiled train/test windows.  For split $j$,
\begin{align}
\mathcal D^{(j)}_{\mathrm{train}}
&=[t_j-T_{\mathrm{train}},t_j),\\
\mathcal D^{(j)}_{\mathrm{test}}
&=[t_j,t_j+T_{\mathrm{test}}),\\
t_{j+1}&=t_j+T_{\mathrm{step}}.
\end{align}
The default driver values are
\[
T_{\mathrm{train}}=504,
\qquad T_{\mathrm{test}}=126,
\qquad T_{\mathrm{step}}=126.
\]
Thus, under the default, OOS windows tile without overlap.

\subsection{Training-Window Candidate Score}

Candidate pairs are ranked on the training window using
\begin{align}
J_{\mathrm{train}}(k)
={}&\mathrm{SR}^{\mathrm{net}}(k)
-0.5\,\mathrm{DD}(k)
-0.5\,\mathrm{TO}(k)
\nonumber\\
&-0.5\,\mathrm{VTE}(k)
-1.0\,\mathrm{QLIKE}(k),
\label{eq:app_train_score}
\end{align}
with
\begin{equation}
k_{\mathrm{train}}^\star
=\arg\max_{k\in\mathcal P}J_{\mathrm{train}}(k).
\end{equation}
Training-window 80th percentiles of estimator loss, turnover, and
volatility-tracking error are also used to define constraints.  Pair
objects are reconstructed before OOS evaluation, so fitted
estimator/controller state does not cross the train/test boundary.

\subsection{Headline S\&P 500 Protocol}

The reported S\&P 500 results use the precomputed-pair protocol.
Component ablations use one contiguous OOS period from 2023-02-10 to
2026-02-10. The model sweep uses 504 training days and a 252-day test/step
configuration, with a single OOS window and unfrozen OOS metrics. Thus, the
headline S\&P results are a contiguous OOS evaluation with train-window
preselection, not an average over repeated walk-forward folds.

\subsection{Multi-Asset Protocol and First-Year Tuning}

The multi-asset evaluation starts on 2024-02-09, with history and metric
construction beginning on 2023-02-10. The reported configuration restricts the
252-observation tuning block to information available before the evaluated OOS
segment; evaluation outcomes are not fed back into the tuning grid. The baseline
grid is
\[
h_{\mathrm{EWMA}}\in\{10,21,42\},
\]
\[
w_{\max}^{\mathrm{asset}}\in\{0.35,0.45,0.60\},
\qquad
G_{\max}\in\{1.0,1.25,1.5\}.
\]
The tuning score is minimized:
\begin{equation}
J_{\mathrm{tune}}
=\overline e_{\mathrm{vol}}
+0.02\,\overline\tau
-0.01\,\mathrm{SR}.
\end{equation}

\section{Information Timing and No-Lookahead Audit}
\label{app:noleak}

\begin{table*}[t]
\centering
\caption{Decision-time information set.}
\label{tab:app_timing}
\scriptsize
\setlength{\tabcolsep}{3pt}
\begin{tabular}{llll}
\toprule
\textbf{Quantity} & \textbf{Window / Source}
& \textbf{Includes time $t$?} & \textbf{Use} \\
\midrule
$\widehat\sigma_t$ & $r^c_{t-W:t-1}$ & No & estimator input \\
Market features & $r^c_{t-W:t-1}$ & No & router state \\
Controller state update & $r_t^c$ & Yes & affects action applied to $t+1$ \\
Pair performance history & past OOS history, appended after $t$ & No & router context \\
Portfolio weight & formed at $t$ & --- & applied to $r_{t+1}$ \\
Realized P\&L & raw $r_{t+1}$ & future to decision & outcome \\
Transaction cost & turnover generated at $t$ & --- & deducted from outcome \\
\bottomrule
\end{tabular}
\end{table*}

The controller may update internal state using the realized return at the
close of $t$, but the resulting action is applied only to $r_{t+1}$.  By
contrast, the estimator and market-state feature window ends at $t-1$.
This estimator/controller timing asymmetry is disclosed explicitly.

The precomputed regime-series generator also enforces a strictly prior
feature window.  For prediction date $t$,
\[
\mathcal W_t=\{r^c_{t-W},\ldots,r^c_{t-1}\}.
\]
Each generated row stores both the prediction date and the last feature
date, making the lag auditable.

The multi-asset implementation contains a dedicated no-leak check for
pair-history consumption, requiring router context to stop at the previous
available date. Table~\ref{tab:app_timing} states the timing convention used
throughout the reported analysis. Auxiliary implementation paths that are
not used by a reported experiment are outside the scope of this timing
statement.

\section{Metric Definitions}
\label{app:metrics}

Let $\{r_t^p\}_{t=1}^T$ denote net strategy returns and $A$ the annualization
factor.  The multi-asset geometric annualized return is
\begin{equation}
R_{\mathrm{ann}}
=\left(\frac{V_T}{V_0}\right)^{A/T}-1.
\end{equation}
Annualized volatility is
\begin{equation}
\sigma_{\mathrm{ann}}
=\operatorname{Std}(r_t^p)\sqrt{A}.
\end{equation}
The implementation-level Sharpe ratio is
\begin{equation}
\mathrm{SR}
=\frac{\overline r^p}{s(r^p)}\sqrt{A},
\label{eq:app_sharpe}
\end{equation}
without subtracting a risk-free rate.  This convention should be noted in
the multi-asset setting because strategy returns may include a risk-free
cash leg.

For the implementation-level Sortino denominator, define downside returns
$d_t=\min(r_t^p,0)$ and
\begin{equation}
\sigma_{\mathrm{down}}
=
\sqrt{A}
\sqrt{\frac{1}{T}\sum_{t=1}^{T}d_t^2}.
\end{equation}
The corresponding annualized Sortino ratio is
\begin{equation}
\mathrm{Sortino}
=
\frac{A\,\overline r^p}{\sigma_{\mathrm{down}}}.
\label{eq:app_sortino}
\end{equation}

Define the running peak $M_t=\max_{j\le t}V_j$.  Maximum drawdown is
\begin{equation}
\mathrm{MDD}
=-\min_t\left(\frac{V_t}{M_t}-1\right).
\end{equation}
For the empirical fifth percentile $q_{0.05}$,
\begin{equation}
\mathrm{CVaR}_{0.95}
=-\mathbb E\left[r_t^p\mid r_t^p\le q_{0.05}\right].
\end{equation}
The implementation Calmar ratio is
\begin{equation}
\mathrm{Calmar}
=\frac{A\overline r^p}{\mathrm{MDD}}.
\end{equation}
Single-asset and multi-asset turnover are, respectively,
\begin{align}
\tau_t^{\mathrm{single}}&=|w_t-w_{t-1}|,\\
\tau_t^{\mathrm{multi}}&=\sum_i|w_{t,i}-w_{t-1,i}|.
\end{align}

Some single- and multi-asset metric paths use different
standard-deviation degrees-of-freedom conventions. Comparisons in the main
paper are therefore interpreted \emph{within each setting}, where all methods
share the same metric implementation, rather than by pooling absolute metric
levels across settings.

\section{Evaluation Scope and Statistical Reporting}
\label{app:reliability}

The main paper reports the common headline metrics available for all four
settings: annualized return, annualized volatility, Sharpe ratio, maximum
drawdown, and CVaR. Additional implementation outputs include Sortino,
turnover, switch counts, selected-pair identities, and volatility-tracking
diagnostics for subsets of experiments.

Headline portfolio statistics in the current study are reported as point
estimates. The manuscript does not attach block-bootstrap confidence
intervals or formal multiple-comparison-adjusted significance tests to the
main table. Accordingly, statements about outperformance are descriptive of
the reported held-out backtests rather than claims of population-level
statistical dominance. For hosted LLM inference, temperature is fixed at
zero, but provider-side nondeterminism can still produce run-to-run variation.

\section{Component Ablations}
\label{app:ablations}

The ablation study intervenes on three parts of the routing process:
state context, candidate/reference evaluation, and adaptive routing. The
non-ablated configuration is reported as \textbf{Full VolRouter}. The main
figure reports Sharpe under the ablation driver; its error bars are properties
of that driver and are not confidence intervals for the headline table.

A material implementation distinction is that ``baseline reference'' has
setting-specific meaning: in S\&P 500 it refers to benchmark-comparison
context, whereas in Multi-Asset it can refer to the deterministic
train-window champion. These interventions are therefore not interpreted as
identical cross-setting ablations.

\section{Router Backbone Study}
\label{app:model_study}

Figure~\ref{fig:router_models} compares the four router backbones labeled in
the figure across S\&P 500, Bitcoin, USDT, and Multi-Asset. The comparison is
intended to test whether routing behavior is specific to a single backbone,
not to establish a universal ranking of language models. In the S\&P sweep,
the regime series is precomputed and frozen across backbones. That panel
therefore compares the downstream switch-review and pair-selection layers,
not the complete state-inference pipeline.

\section{Candidate-Pool and Sensitivity Diagnostics}
\label{app:robustness}

The S\&P candidate-pool diagnostic uses pool sizes
\[
27,42,57,72,87,102,117.
\]
The Multi-Asset reduction driver follows a different procedure: it removes
15 pairs per round using seed 20260703, reranks surviving pairs using
pre-test data, and truncates the candidate list to at most 30 pairs. Thus,
``pool size'' and the number of candidates actually exposed to the Router
are distinct quantities in that driver.

The four sensitivity panels in Figure~\ref{fig:router_sensitivity} vary
switching sensitivity, transaction cost, target volatility, and candidate
pool size one factor at a time. The dashed line is the canonical base
configuration and is not necessarily included as a plotted sweep point. In
particular, the displayed switching sweep reports very low, low, high, and
very high settings around a separate normal/base reference. The transaction
cost values shown in Figure~\ref{fig:router_sensitivity} are specific to that
routing sensitivity run and should not be conflated with other
controller-level friction grids in the codebase.

\section{Switching, Turnover, and Execution Cost}
\label{app:switching}

Define
\begin{equation}
N_{\mathrm{switch}}
=\sum_{t=1}^{T}\mathbf 1[k_t\neq k_{t-1}],
\qquad
\mathrm{SwitchRate}=\frac{N_{\mathrm{switch}}}{T}.
\end{equation}
If the active-pair sequence forms $J$ holding segments with lengths
$\ell_1,\ldots,\ell_J$, mean dwell time is
\begin{equation}
\overline\ell=\frac{1}{J}\sum_{j=1}^{J}\ell_j.
\end{equation}

Portfolio turnover costs are charged through changes in the executed
portfolio weights. In the reported Multi-Asset ablation, pool-reduction, and
sensitivity drivers, the additional monetary penalty attached solely to a
change in pair identity is zero. Switching is instead regularized by the
hold/switch gate, minimum-hold logic, and persistence guidance. This is why
the main-text net-return equation includes portfolio turnover cost but not a
uniform inter-pair switching fee.

\section{Reproducibility Notes}
\label{app:reproducibility}

Structured router outputs are validated before execution, and failed or
invalid outputs follow the deterministic fallback logic described in
\AppendixRef{app:prompt_routing_format}. Several machine-learning estimators
use \texttt{random\_state}=42, while the pool-reduction diagnostic uses seed
20260703. Hosted LLM generation is not fully seed-controlled even at
temperature zero; the backbone analysis is therefore interpreted as an
empirical comparison of the reported runs rather than a deterministic model
ranking.

\end{document}